%% file: manuscript.tex
\documentclass[journal]{IEEEtran}

\ifCLASSINFOpdf
\else
\fi

\usepackage{cite}
\usepackage{times}
\usepackage{graphicx}
\usepackage{amsmath}
\usepackage{algorithm}
\usepackage[noend]{algorithmic}
\usepackage{subfig}
\usepackage{stfloats}
\usepackage{caption}
\usepackage{color}
\usepackage{fancyvrb}
\usepackage{amsmath}
\usepackage{amssymb}
\usepackage{makeidx}
\usepackage{mdwmath}
\usepackage{mdwtab}
\usepackage{moreverb}
\usepackage{fancyvrb}
\usepackage{etoolbox}
\usepackage{mathtools}
\usepackage{mathptmx}
\usepackage{multirow}
\usepackage{color}
\usepackage{url}
\usepackage{tabularx}

\usepackage{hyperref}
\usepackage{booktabs}
\usepackage{tikz}

\usepackage{titlesec}
\titlespacing\section{0pt}{12pt plus 4pt minus 2pt}{2pt plus 2pt minus 2pt}
\titlespacing\subsection{0pt}{12pt plus 4pt minus 2pt}{2pt plus 2pt minus 2pt}

\usepackage[nolist]{acronym}

\input{Acronyms}

\begin{document}
%


\title{
    \vspace{-0cm} 
    \begin{tikzpicture}[remember picture, overlay]
        \node[anchor=north, yshift=-0.5cm] at (current page.north) {\fbox{\parbox{\textwidth}{\centering\small {\color{red}This version of the paper has been submitted for publication to IEEE and may be removed upon request or copyright issues.}}}};
    \end{tikzpicture}
    \vspace{0cm} 
    QACM: QoS-Aware xApp Conflict Mitigation in Open RAN

}


%
%

\author{Abdul Wadud,~\IEEEmembership{Graduate Student Member,~IEEE,}
Fatemeh Golpayegani,~\IEEEmembership{Senior Member,~IEEE,}
and~Nima Afraz,~\IEEEmembership{Senior Member,~IEEE}
\thanks{Manuscript received on X, 2024; revised on Y, 2024; and accepted on Z, 2024.}
\thanks{Abdul Wadud is with the School of Computer Science, University College Dublin, Ireland and Bangladesh Institute of Governance and Management, Dhaka, Bangladesh}
\thanks{Fatemeh Golpayegani and Nima Afraz are with the School of Computer Science, University College Dublin, Ireland.}}

\maketitle

\begin{abstract}
The advent of Open \ac{RAN} has revolutionized the field of \ac{RAN} by introducing elements of native support of intelligence and openness into the next generation of mobile network infrastructure. Open \ac{RAN} paves the way for standardized interfaces and enables the integration of network applications from diverse vendors, thereby enhancing network management flexibility. However, control decision conflicts occur when components from different vendors are deployed together. This article provides an overview of various types of conflicts that may occur in Open \ac{RAN}, with a particular focus on intra-component conflict mitigation among \acp{xApp} in the \ac{Near-RT-RIC}. A \ac{QACM} method is proposed that finds the optimal configuration of conflicting parameters while maximizing the number of \acp{xApp} that have their \ac{QoS} requirements met. We compare the performance of the proposed \ac{QACM} method with two benchmark methods for priority and non-priority cases. The results indicate that our proposed method is the most effective in maintaining \ac{QoS} requirements for conflicting \acp{xApp}. \footnote{This paper is an extension of \cite{wadud2023conflict}, presented at IEEE CPSCom 2023.}
\end{abstract}

\begin{IEEEkeywords}
Open \ac{RAN}, conflict mitigation, \ac{QoS}, \ac{xApp}, \ac{Near-RT-RIC}.
\end{IEEEkeywords}


%
\IEEEpeerreviewmaketitle

\section{Introduction}
%
%
%
%
\IEEEPARstart{T}{he} wireless \ac{RAN} has transformed over the past few decades as we have witnessed remarkable progress in wireless communication technologies. Many novel \ac{RAN} concepts other than the Traditional Radio Access Network or \ac{D-RAN} have been introduced to support the more diverse and stringent network requirements of \ac{5G} and beyond communication systems including \ac{C-RAN}, \ac{V-RAN}, \ac{SD-RAN}, and Open Radio Access Network.
Open \ac{RAN} is considered one of the most promising because of its disaggregated, virtualized, and vendor-neutral architecture. It provides a native framework to support \ac{AI}/\ac{ML}-based control applications that enhance the network and resource management for the \acp{MNO}. Contrarily, the other versions of \ac{RAN} architectures are typically deployed by a single vendor that works as a black-box solution, which risks vendor lock-in for the \acp{MNO}. Therefore, Open \ac{RAN} is becoming more crucial for the \acp{MNO}. 


The O-RAN ALLIANCE is a worldwide community of \acp{MNO}, vendors, and research \& academic institutions that envisions the future as intelligent, open, virtualized, and fully interoperable Open \ac{RAN}. Open \ac{RAN} architecture has several significant challenges to overcome, as it is still less mature compared to other RAN architectures. Security \& trust issues and interoperability are two of the major challenges. As the disaggregated Open \ac{RAN} architecture uses multiple splits between components of the \ac{RAN} protocol stacks, it opens many loopholes for attackers. A significant amount of studies are required to enhance trust and security in the Open \ac{RAN}.

Interoperability remains another major challenge in Open RAN. When components from different vendors are assembled together, they should operate seamlessly, without any conflict or with minimal conflict. Conflicts negatively impact network performance and degrade the \acp{KPI}. There is neither a standardized conflict mitigation framework, nor is it defined how components from various vendors may coordinate, according to the O-RAN ALLIANCE. Therefore, in this study, we focus on post-action \ac{QoS}-aware conflict mitigation within the \ac{RIC} to reduce the negative impact of conflicts. To elaborate, the \ac{RIC} is responsible for network control within the open \ac{RAN} architecture. It allows various vendors to deploy control applications aimed at specific network objectives, such as resource allocation, energy saving, mobility load balancing, and more {\cite{polese2022understanding}}. Tasks that are not time-sensitive, taking longer than $1s$ to complete, are managed by \acp{rApp} within the \ac{Non-RT-RIC}. In contrast, tasks demanding completion in a time frame from $10ms$ to $\leq 1s$ are performed in \acp{xApp} in the \ac{Near-RT-RIC}.
The \acp{xApp} and \acp{rApp} within the \ac{RIC} oversee network operations and management. An example is an \ac{rApp} that enhances network efficiency and minimizes delays in \ac{V2X} communication by optimizing the allocation of radio resources \cite{rimedo_labs}. In a similar manner, an \ac{xApp} can optimize \ac{QoS} for a user group by efficiently managing radio resources and dispatching targeted control signals to the \ac{RAN} infrastructure \cite{b3}. There is a risk of adverse interactions affecting performance \cite{ ric_oran_alliance} given that these \acp{xApp} and \acp{rApp} are supplied by different vendors and operate on shared resources during network activities. Such interactions are called conflicts that must be identified and resolved to prevent significant declines in system performance.

The impact of conflict at the \ac{xApp} level is significant. Recent studies \cite{adamczyk2023conflict, zhang2022team, del2024pacifista} demonstrate, degradation in \ac{RAN} \acp{KPI} when \acp{xApp} operate independently without any conflict mitigation model. For instance, the authors in \cite{adamczyk2023conflict} examine \ac{MLB} and \ac{MRO} \acp{xApp}, showing that a simple prioritization-based conflict resolution method can improve handover \acp{KPI} compared to scenarios without conflict mitigation. Similarly, \cite{zhang2022team} reveals that resource and power allocation \acp{xApp} employing a team-learning based conflict mitigation method significantly outperforms independent operations in terms of throughput and packet drop rate (PDR). Finally, research by \cite{del2024pacifista} shows that throughput drops by approximately 50\% when \ac{ES} and throughput maximization (TM) \acp{xApp} operate together in a \ac{RAN} slice due to conflicting configurations, reinforcing the need for research into \ac{xApp} conflict mitigation in Open \ac{RAN}.

The study in \cite{adamczyk2023conflict} focuses solely on conflict detection and mitigation frameworks, applying the \ac{xApp} prioritization method exclusively to indirect conflicts. In contrast, \cite{zhang2022team} introduces a \ac{DRL}-based team learning method that necessitates data sharing among \acp{xApp} and incurs high computational costs, limiting scalability with an increasing number of \acp{xApp}. Meanwhile, \cite{del2024pacifista} presents a threshold-based mitigation approach, setting tolerance thresholds for each \ac{xApp} to gauge conflict severity. If two \acp{xApp} are producing conflict that is above the severity threshold, the \ac{MNO} should not deploy them together to avoid conflict. However, it lacks discussion on how altering conflicting parameter values could mitigate this severity when it is above the threshold. These limitations motivate us to investigate a more efficient conflict mitigation method that is computationally less intensive, does not require data sharing among \acp{xApp}, can effectively mitigate conflict severity in \ac{RAN}, and can ensure the \ac{QoS} of the network (see more in Section~\ref{sec:problem_back}).

In this article, we propose the \ac{QACM} method to address various types of intra-component conflicts in the \ac{Near-RT-RIC} while ensuring the individual \ac{QoS} requirements of conflicting \acp{xApp}. This proposed method extends our previous research \cite{wadud2023conflict}, where we introduced two game-theory-based \acp{CMC}, namely \ac{NSWF} and \ac{EG} solutions. The \ac{NSWF} is applied in non-priority scenarios where each \ac{xApp} in conflict has equal preference, while \ac{EG} is utilized in priority settings, allowing \acp{xApp} to have varying preferences set by the \ac{MNO}. However, these methods do not account for the \ac{QoS} benchmarks of each associated \ac{KPI} of the involved \acp{xApp} during conflict mitigation. Hence, many \acp{xApp} fall short of their \ac{QoS} requirements with the provided solutions. Thus, this article proposes the \ac{QACM} framework, considering the \ac{QoS} benchmarks of conflicting \acp{xApp}. We benchmark the \ac{NSWF} and \ac{EG} solutions against the proposed approach and compare their performances in Section~\ref{sec:case_study}.

The concept and architecture of the \ac{CMS} are adopted from \cite{adamczyk2023conflict}. While our research mainly focuses on the \ac{CMC} component of the \ac{CMS}, the study in \cite{adamczyk2023conflict} concentrated on the \ac{CDC}. To the best of our knowledge the proposed \ac{QACM} method is the first of its kind for \ac{RAN} conflict mitigation.

We formulate the \ac{QACM} method as an optimization problem and heuristic algorithm in Section~\ref{sec:proposed} and provide an in-depth discussion on the taxonomy of conflict in Open \ac{RAN} in Section~\ref{sec:background}. Furthermore, we present an example model with five stochastic \acp{xApp} to illustrate different types of intra-component conflicts in the \ac{Near-RT-RIC} and to theoretically analyze the performance of the proposed \ac{QACM} method compared to benchmarks. The case study demonstrates that the proposed \ac{QACM} method outperforms benchmarks in maintaining the \ac{QoS} threshold of involved \acp{xApp}. We conduct four different case studies that cover conflicting cases considering two or more involved \acp{xApp} and methods for handling different types of conflicts both separately and together. This paper makes several significant contributions to the field of Open \ac{RAN} conflict mitigation, particularly in the context of \ac{Near-RT-RIC}:

\begin{itemize}
    \item We provide a comprehensive taxonomy of conflicts in Open \ac{RAN} and discuss the specific challenges and methodologies for mitigating intra-component conflicts within the \ac{Near-RT-RIC}.
    \item we demonstrate different conflicts using an example model with five stochastic \acp{xApp}.
    \item We provide a comprehensive overview of the \ac{CMS} and its components for mitigating intra-component conflicts within the \ac{Near-RT-RIC}.
    \item We propose the \ac{QACM} method to specifically address various types of conflicts among \acp{xApp} while ensuring the \ac{QoS} requirements of each \ac{xApp} are met.
    \item A novel optimization problem and heuristic algorithm for the \ac{QACM} focusing on improving the computational efficiency and effectiveness of conflict mitigation.
    \item Four case studies alongside a simulation study of direct conflict in a RAN environment to empirically validate the effectiveness of the \ac{QACM} method, showing its superiority in maintaining the \ac{QoS} thresholds of involved \acp{xApp} over traditional methods.
    \item Finally, the research offers insights highlighting the importance and practical implications of deploying \acp{xApp} from diverse vendors in a standardized interface environment with a mitigation method.
\end{itemize}

These contributions collectively advance the state-of-the-art conflict mitigation strategies within the Open \ac{RAN} ecosystem, offering scalable, efficient, and effective solutions for real-world deployment challenges. The remainder of the article is organized as follows: Section~\ref{sec:background} covers the background of conflict with a state-of-the-art literature review. Section~\ref{sec:system_model} discusses the system model for the proposed \ac{QACM} method within the \ac{CMS} framework. Section~\ref{sec:cms_framework} discusses the conflict management framework within the \ac{Near-RT-RIC} architecture of Open \ac{RAN}. Section~\ref{sec:bench} discusses the benchmark methods that are used to compare the performance with the proposed \ac{QACM} method.  Section~\ref{sec:proposed} elaborates on the proposed \ac{QACM} method and its prerequisites. Section~\ref{sec:case_study} analyzes the performance comparison between the proposed and benchmark methods through case studies. Section~\ref{sec:application} discusses the real-world application of the proposed method. Section~\ref{sec:limit_fut} addresses the limitations and future work. Finally, the paper concludes in Section~\ref{sec:conclusion}.

\section{Background}
\label{sec:background}

\subsection{Conflicts in Open RAN}
\label{sec:conflicts_in_oran}
In a conventional \ac{RAN} setup with a single vendor, the vendor typically managed and resolved any conflicts within their own architecture. As the sole provider of the \ac{RAN} system, they oversaw the design, setup, and fine-tuning of the network. They address any issues or incompatibilities within their exclusive ecosystem, leading to a more simple prevention and resolution strategy. However, the advent of Open \ac{RAN} has changed this dynamic. This new network structure supports the incorporation of hardware and software from multiple vendors. While this enhances interoperability and adaptability, it also brings about possible discrepancies among the different components. Each vendor might employ distinct methods, enhancements, or settings, which can cause disputes when merging their technologies into the Open \ac{RAN} environment and adversely affect the \ac{RAN}'s efficiency. 

Consequently, these conflicts need to be identified and managed through appropriate network management \cite{ric_oran_alliance}. 
\subsubsection{Taxonomy of Conflicts in Open RAN}
\label{types_conf}
\begin{figure}[!ht]
\centering
\vspace{-0.1in}
\includegraphics[scale = 0.42]{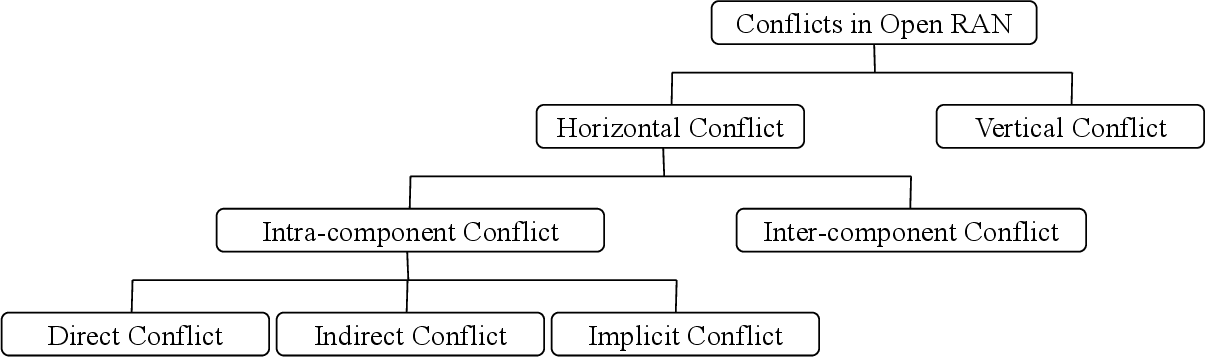}
\caption{Taxonomy of potential conflicts in Open \ac{RAN} \cite{wadud2023conflict}.}
\label{fig:tax}
\vspace{-0.1in}
\end{figure}

Recent studies \cite{adamczyk2023conflict, wadud2023conflict} have indicated that control decision conflicts in Open \ac{RAN} architecture can manifest at various levels. These conflicts within Open \ac{RAN} are typically divided into horizontal and vertical types, as depicted in Fig.~\ref{fig:tax}.

\par
\begin{figure}[!ht]
 \centering
	\includegraphics[scale=0.6]{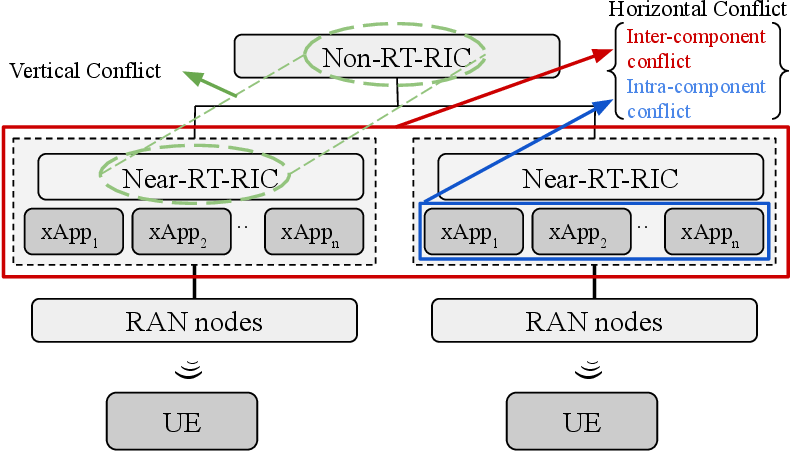}
	\caption{Potential conflicting areas in Open \ac{RAN} \cite{wadud2023conflict}.}
	\label{fig:area_conf}
 \vspace{-0.1in}
\end{figure}

 A vertical conflict emerges between components at different layers of the Open \ac{RAN} hierarchy. For instance, a conflict between a \ac{Near-RT-RIC} and a \ac{Non-RT-RIC}, as shown in Fig.~\ref{fig:area_conf}, is identified as a vertical conflict. Conversely, a horizontal conflict arises among components at the same hierarchical level. An example is a dispute between two \acp{xApp} within a single \ac{Near-RT-RIC} or among adjacent \acp{Near-RT-RIC}, which is classified as a horizontal conflict (refer to Fig.~\ref{fig:area_conf}). Within a \ac{Near-RT-RIC}, conflicts among \acp{xApp} are termed intra-component conflicts, while those among \acp{xApp} from neighboring \acp{Near-RT-RIC} are called inter-component conflicts (see Fig.~\ref{fig:area_conf}). Furthermore, intra-component conflicts can be broken down into direct, indirect, and implicit types. In this paper, we present a strategy to mitigate intra-component conflicts among \acp{xApp} in the \ac{Near-RT-RIC}.
Within the \ac{Near-RT-RIC}, autonomous \acp{xApp} aiming for various optimization objectives can inadvertently create conflicting configurations by altering or affecting the same network parameter \cite{ric_oran_alliance}. These are recognized as intra-component conflicts. Resolving direct, indirect, and implicit conflicts is challenging as the \acp{xApp} involved are often developed and provided by different vendors and typically do not share information with each other \cite{zhang2022team}.

Direct conflicts are easily identifiable by the internal conflict mitigation controller. A single \acp{xApp} or a pair of \acp{xApp} might request configurations that clash with the existing setup. Furthermore, several \acp{xApp} might propose different values for the same parameter, leading to an evident direct conflict. The conflict mitigation controller evaluates these requests and determines which one should take precedence. This approach is referred to as pre-action resolution \cite{ric_oran_alliance}. However, simply favoring one request over another is not always the best solution. A preferable strategy is to find an optimal configuration that reconciles the conflicting parameters, promoting fairness and aligning with the network's collective optimization objectives.
In contrast, indirect and implicit conflicts are less obvious. An indirect conflict arises when one \ac{xApp}'s parameter adjustment inadvertently affects the operational area of another \acp{xApp}. For instance, separate \acp{xApp} controlling cell individual offset (CIO) and antenna tilts might influence the handover boundary. A change by an \ac{xApp} managing remote electrical tilts (RET) or antenna tilts can, therefore, indirectly alter the performance of a CIO-focused \acp{xApp}. Addressing this type of conflict involves post-action analysis to determine an optimal value for the contentious parameter \cite{ric_oran_alliance}. Implicit conflicts occur when two \acp{xApp}, each optimizing their respective targets, inadvertently degrade each other's performance. An \ac{xApp} aimed at ensuring \ac{QoS} for a set of users and another focused on minimizing handovers could, for instance, interfere with one another in subtle ways. Detecting and resolving this conflict is particularly challenging \cite{ric_oran_alliance}. 
In this article, we introduce a conflict mitigation component designed to address all intra-component conflicts among \acp{xApp} in the \ac{Near-RT-RIC}.
\vspace{-10pt}
\subsection{Problem Background}
\label{sec:problem_back}
The \ac{Near-RT-RIC} is the core of control and optimization in Open \ac{RAN}. \acp{xApp} are the main components of the \ac{Near-RT-RIC}. According to the O-RAN Alliance's Open \ac{RAN} architecture, it opens the network for smaller vendors to participate in the development of \ac{RAN} components so that the over-dependency of \acp{MNO} on a handful of vendors can be alleviated. When these \acp{xApp} from various vendors are deployed in the \ac{Near-RT-RIC}, the possibility of having conflicting configurations among them is highly likely or certain as they share the same \ac{RAN} resources \cite{wadud2023conflict, zhang2022team, adamczyk2023conflict, adamczyk2023challenges}. The common approach to deal with these conflicts is to develop a combined \ac{xApp} that performs multiple objectives, for instance- traffic steering \ac{xApp} \cite{dryjanski2021toward}, or enabling real-time data-sharing between these conflicting \acp{xApp} for joint decision making using \ac{MARL} or other machine learning techniques \cite{zhang2022team}. 
 From the \ac{MNO}'s perspective, the former may loop us back to the possibility of vendor-lock-in as it acts like a black-box of multiple tasks combined together. The latter will require real-time data sharing and management that adds excessive computational overheads in the latency sensitive \ac{Near-RT-RIC}. Also, it fails when vendors are not interested in direct data sharing of their \ac{xApp}'s with others. In our previously published paper in \cite{wadud2023conflict}, we proposed a game theoretic approach that can satisfy the low latency scenario considering the following few assumptions:
\begin{itemize}
    \item Each \ac{xApp} possesses the capability to independently learn, predict, and make optimal decisions based on the changing network state.
    \item \acp{xApp} are capable of operating autonomously without intercommunication.
    \item Provided by various vendors, \acp{xApp} function independently without forming groups or collective actions among themselves.
    \item All \acp{xApp} utilize shared resources, leading to inherent conflicts of interest among them.
\end{itemize}


\par In our previous study \cite{wadud2023conflict}, the \ac{NSWF} is utilized to optimize a parameter that several \acp{xApp} conflict over. This approach primarily aims to maximize the system's overall utility as shown in the equation~\ref{eq:nswf}. However, it does not consider the \ac{QoS} requirements of individual \acp{xApp}, potentially causing them to consistently fall below their \ac{QoS} targets. To address this, we propose a \ac{QoS}-aware conflict mitigation model considering all four aforementioned assumptions. This model focuses on ensuring that each \ac{xApp} closely meets its individual \ac{QoS} requirements as well as the overall network performance.

\subsection{Related Research}
\label{sec:related_research}
Open \ac{RAN} is a relatively new concept, and its conflict management aspects have not been thoroughly explored yet. However, there has been substantial research on conflict management within \acp{SON}. The idea of \acp{xApp} bears similarities to the concept of centralized \acp{SF}, as the \ac{RIC} is seen as an evolution of the \ac{SON} \cite{Mavenir_RIC}. \ac{SON} was introduced to simplify and automate the deployment and optimization of cellular \acp{RAN} by removing the need for manual network element configuration. Consequently, it lowers the operational costs for mobile operators, enhances the \ac{OpEx}-to-revenue ratio, and postpones unnecessary \ac{CapEx}.
As the telecom industry moves towards open interfaces, virtualization, and software-centric networking, the \ac{SON} ecosystem is gradually shifting from traditional \ac{D-SON} and \ac{C-SON} models to a framework based on open standards as \acp{xApp} and \acp{rApp}. This transition closely aligns with \ac{RAN} programmability, fostering advanced automation and intelligent control through the \ac{RIC}. The \ac{RIC}, \ac{xApp}, and \ac{rApp} are equipped to support both near real-time \ac{D-SON} and non-real-time \ac{C-SON} functionalities, meeting the \ac{RAN} automation and optimization requirements effectively \cite{SNS_SON}. Therefore, conflict mitigation in \ac{SON} is relevant and essential to studying the conflict in Open \ac{RAN}. 
Conflicts between \ac{MLB} and \ac{MRO} \acp{SF} are frequently studied within the context of \ac{SON} \cite{liu2010conflict, mu2014conflict, huang2018conflict}. The research in \cite{liu2010conflict} addresses the conflict between \ac{MLB} and \ac{MRO} by restricting the \ac{CIO} parameter's range as determined by the \ac{MLB}. The study conducted in \cite{huang2018conflict} explored finding an optimal value for the \ac{CIO} to enhance the mitigation of conflicts between these two \acp{SF}. Munoz et al., in \cite{mu2014conflict} introduced a coordination algorithm aimed at resolving the dispute between these \acp{SF} by adhering to pre-defined threshold values for \ac{HOR} and \ac{CBR}. A superior \ac{HOR} above the threshold suggests improved performance for the \ac{MRO}, while a lower \ac{CBR} under the threshold signifies reduced cell-site congestion and thus better performance for the \ac{MLB}. 
The study in \cite{anubhab2020conflict} introduced a game-theoretic approach to mitigating conflicts among \acp{CF} in \acp{CAN}. \acp{CAN} represent an advanced version of \acp{SON} that employ machine learning and artificial intelligence to analyze network data and construct models depicting network behavior. The researchers developed a machine learning-based regression model for each \ac{CF} using data gathered from the network. This data was collected for every \ac{CF} through a simulated experimental setup that replicates real network conditions, encompassing all \acp{ICP} and \acp{KPI} associated with each \ac{CF}. The \ac{NSWF} was applied within three distinct conflict models to determine the optimal values for the conflicting \acp{ICP} while enhancing the overall utility of the network. This research was further expanded in \cite{banerjee2021toward} with actual network data obtained from a network simulator, considering both priority and non-priority scenarios for the \acp{CF}. 

A recent study in \cite{zhang2022team} adopts a reinforcement learning based Deep Q-learning model for cooperative learning between power allocation and resource allocation \acp{xApp} in Open \ac{RAN}. The results showed higher throughput and lower packet drop rate while considering the team learning approach compared to the non-team learning approach. This approach demonstrated a new solution to resolve the conflicting problem that might be viable to adapt in Open \ac{RAN}, but seems hard to excel for a higher number of \acp{xApp} because of the complexity of joint decision-making for multiple participating \acp{xApp} using the demonstrated framework. Moreover, \acp{xApp} should be designed and developed with this proposed framework in mind, otherwise, the solution cannot be adopted. The research by \cite{del2024pacifista} introduces a mitigation method based on setting tolerance thresholds for each \ac{xApp}, which helps measure the severity of conflicts. If the conflict between any two \acp{xApp} exceeds this severity threshold, they should not be deployed simultaneously by the \ac{MNO} to prevent further conflicts. However, this study does not explore how adjustments to conflicting parameter values could alleviate such severity once it surpasses the threshold. Table~\ref{tab:conflict_mitigation} provides a comparative summary of the proposed \ac{QACM} method alongside the three state-of-the-art methods discussed.

Researchers have developed a conflict detection and resolution framework, as outlined in \cite{adamczyk2023conflict}, which includes components for direct, indirect, and implicit conflict detection to effectively identify them within the \ac{Near-RT-RIC}. The research primarily concentrates on the detection phase, while employing a simplistic priority-based system for conflict mitigation. Their findings demonstrated positive results, particularly with \ac{MLB} and \ac{MRO} \acp{xApp}. Nonetheless, modifications to the current Open \ac{RAN} architecture are necessary to integrate this solution. \cite{adamczyk2023challenges} explores the potential challenges of mitigating various types of conflicts and discusses control loops for three different types of conflict mitigation approaches, including- preventive conflict mitigation, conflict detection and resolution, and supervision \& adaptation. The preventive conflict mitigation approach suggests pre-deployment assessment of \acp{xApp} and \ac{rApp} in a digital-twin environment to detect potential conflicts and analyze their impacts on the network before deploying them to the actual \ac{RIC}. The conflict detection and resolution approach is a post-action method in the live network that detects and resolves any type of conflict in near-real-time. The last of these three envisions to have a conflict supervisor component on top of the conflict detection and resolution framework that provides closed-loop monitoring and reconfiguration while mitigating conflict in the \ac{RIC}. 
\begin{table*}[ht]
\centering
\caption{Comparison of this study with articles focusing on conflict mitigation methods in Open RAN}
\label{tab:conflict_mitigation}
\begin{scriptsize} 
\begin{tabular}{|>{\color{black}}p{3.1cm}|p{1.9cm}|p{2cm}|p{1.5cm}|p{1.5cm}|p{1.5cm}|p{1.5cm}|p{1.5cm}|} 
\hline
\textbf{Article} & \textbf{Method} & \textbf{Conflicts Covered} & \textbf{Ensures QoS} & \textbf{Parameter Control} & \textbf{Dynamic Mitigation} & \textbf{Data Sharing} & \textbf{Scalability} \\ \hline
Adamczyk et al. (2023) \cite{adamczyk2023conflict, adamczyk2023challenges} & xApp prioritization & Indirect & \texttimes & \texttimes & \checkmark & \texttimes & Low \\ \hline
Zhang et al. (2022) \cite{zhang2022team} & Team learning & Indirect & \texttimes & \texttimes & \checkmark & \checkmark & Moderate \\ \hline
Prever et al. (2024) \cite{del2024pacifista} & Severity threshold & Direct, indirect and implicit & \texttimes & \texttimes & \texttimes & \texttimes & High \\ \hline
This work & QACM method & Direct, indirect and implicit & \checkmark & \checkmark & \checkmark & \texttimes & High \\ \hline
\end{tabular}
\end{scriptsize}
\vspace{-15pt}
\end{table*}
In our earlier work presented in \cite{wadud2023conflict}, which is a post-action conflict detection and mitigation framework, we primarily concentrated on the conflict mitigation component, assuming that the Conflict Detection (CD) component, with the support of the Performance Monitoring (PMon) component, accurately identifies conflicts. The conflict resolution strategy we adopted utilized two game-theoretic methods: \ac{NSWF} and \ac{EG} solution, for priority-based and non-priority-based scenarios, respectively. However, we observed that both \ac{NSWF} and \ac{EG} occasionally fail to guarantee satisfactory Quality of Service (\ac{QoS}) for the maximum number of conflicting \acp{xApp}. This inadequacy stems from not considering the \ac{QoS} targets of individual \ac{xApp}. We propose a \ac{QoS}-aware conflict mitigation approach designed to ensure that the majority, if not all, \acp{xApp} meet or exceed their specified \ac{QoS} thresholds by identifying an optimal setting for the contentious \ac{ICP}. We discuss this proposed method in Section~\ref{sec:proposed}.

\section{System Model}
\label{sec:system_model}
\subsection{Assumptions and Notations}
Let us assume that there are $n$ \acp{xApp} installed in the \ac{Near-RT-RIC}. The set of \acp{xApp} is denoted by $X$, where $X = \{ x_1, x_2, \ldots, x_n \}$. Each \ac{xApp} $x \in X$ has at least one associated \ac{KPI} $k_j \in K$, where $K$ indicates the set of all \acp{KPI} in the network. If there are multiple \acp{KPI} associated with a single \ac{xApp} $x \in X$, we represent them as $k_{ji}$, where $j$ denotes the index of the \ac{xApp}, and $i$ represents the index of the associated \ac{KPI}. The \acp{KPI} of an \ac{xApp} vary with the change of their associated input control parameters. We define the set of input control parameters as $P$, where $p \in P$. Each \ac{KPI} has an individual \ac{QoS} threshold $q_j \in Q$ to maintain. We define $X^\prime$ and $Q^\prime$ as sets of \acp{xApp} with having a conflict over the conflicting parameter $p_l$ and their associated \acp{KPI}, respectively. Since each \ac{KPI} $k_j \in K$ may have a different unit, we convert them to a scalar unit by the function $U(p)$ and denote by $u \in U$. It is the utility function of $x \in X$ as a function of $p$. Converting the \acp{KPI} to the utility function is one of the most critical challenges of this proposed method, which is discussed in detail in Section~\ref{sec:kpi2uti}. The objective of the proposed method is to select a value for the conflicting parameter $p_l \in P$ within a constrained range that minimizes the distance between the utility of \ac{xApp} $x^\prime \in X^\prime$ and its \ac{QoS} threshold $q^\prime \in Q^\prime$. $s$ is the indicator of whether an \ac{xApp} meets its \ac{QoS} threshold or not for the estimated optimal value of the conflicting parameter $p_l$. $w$ indicates the assigned weight to the conflicting \acp{xApp} by the \ac{CS} \ac{xApp}. The functionalities of \ac{CS} \ac{xApp} are discussed in Section~\ref{sec:cms_framework}. The distance calculated between normalized \ac{QoS} threshold and utility is very small, therefore, we use a constant $\zeta$ to tune the weighted distance. In our numerical analysis, we used $\zeta = 10^3$. All important notations used in this article are summarized in Table~\ref{tab:notation}. 

\begin{figure}[!ht]
 \centering
 \vspace{-0.1in}
	\includegraphics[scale = 0.55]{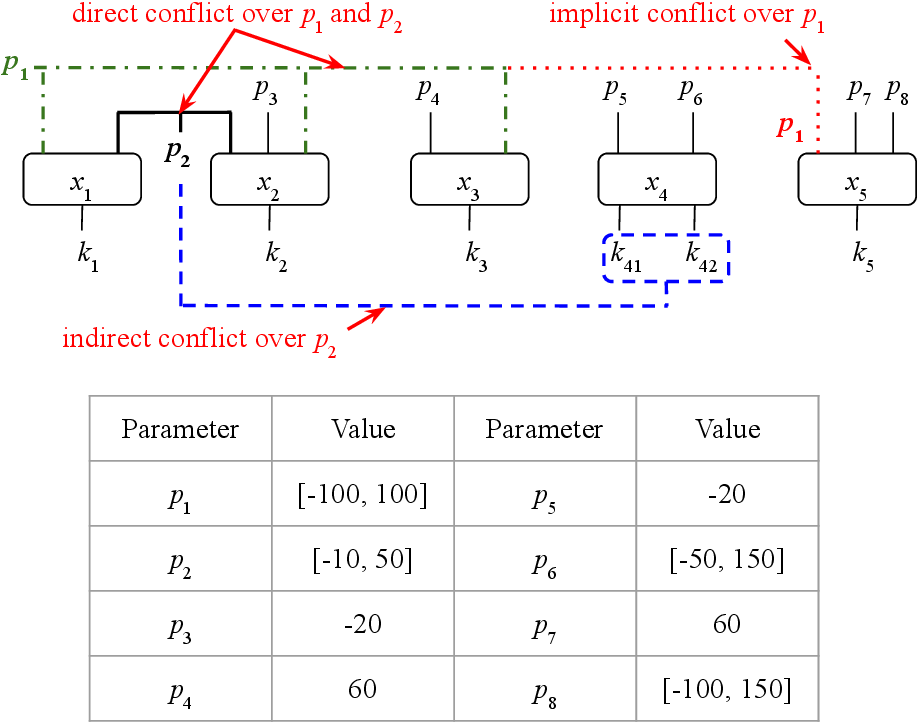}
	\caption{Example model for direct, indirect and implicit conflict.}
	\label{fig:example_model}
 \vspace{-0.1in}
\end{figure}

Fig.~\ref{fig:example_model} depicts the example model used for theoretically analyzing the efficacy of the proposed conflict mitigation method, where five stochastic \acp{xApp} are considered. These \acp{xApp}, installed in the \ac{Near-RT-RIC}, are strategically modeled to encompass all intra-component conflicts. \ac{xApp} $x_1$, $x_2$, and $x_3$ share the \ac{ICP} $p_1$, thus exhibiting direct control decision conflict over $p_1$. Similarly, $x_1$ and $x_2$ have a direct conflict over $p_2$. We examine these two direct conflicts in three distinct scenarios: firstly, addressing direct conflict between two \acp{xApp}; secondly, managing similar conflicts among more than two \acp{xApp}; and lastly, simultaneously resolving two direct conflicts involving multiple \acp{xApp}.

An indirect conflict is modeled considering $p_2$, $p_5$, and $p_6$, which belong to the same parameter group in the database. This means any change in their values impacts \acp{KPI} $k_{41}$ and $k_{42}$. As defined in Section~\ref{sec:conflicts_in_oran}, an indirect conflict arises when one \ac{xApp}'s parameter adjustment inadvertently influences the functional area of another \ac{xApp}. This implies that a change in $p_2$, associated with $x_1$ and $x_2$, that inadvertently affects the \acp{KPI} associated with $x_4$, constitutes an indirect conflict of $x_4$ with both $x_1$ and $x_2$ over $p_2$. For instance- $P_{k_{41}}^G = \{p_2, p_5, p_6\}$ and $P_{k_{42}}^G = \{p_2, p_5, p_6\}$: here, $p_2 \notin I_{x_4}$, but it still affects $k_{41}$ and $k_{42}$. Since $p_2 \in I_{x_1}$ and $I_{x_2}$, we can say there is an indirect conflict of these xApps with $x_4$ over $p_2$. Here, $I_{x_1}$ and $I_{x_2}$ are the set of \acp{ICP} for $x_1$ and $x_2$, respectively.

Lastly, an implicit conflict of $x_5$ with $x_1$, $x_2$, and $x_3$ over $p_1$ is modeled. This indicates that any alteration in $p_1$ inadvertently affects the \ac{KPI} $k_5$ of $x_5$. Although $p_1$ is not directly linked as an \ac{ICP} of $x_5$, it implicitly influences $x_5$'s performance and acts as its implicit input, characterizing implicit conflicts. For instance- the parameter group for $k_5$ is $P_{k_5}^G = \{p_7, p_8\}$: This parameter group indicates that only modifying $p_7$ and $p_8$ should affect $k_5$. However, if during the RAN operation, $k_5$ gets affected by changing $p_1$, then we can say there is an implicit conflict between $x_1, x_2, x_3$ and $x_5$ over $p_1$ since $p_1 \in I_{x_1}$, $I_{x_2}$, and $I_{x_3}$. Including $p_1$ inside $P_{k_5}^G$ changes it to $P_{k_5}^G = \{p_1, p_7, p_8\}$, making it an indirect conflict. Thus, an implicit conflict is a vague form of indirect conflict that can only be detected during RAN operation but can be modeled as an indirect conflict once detected. Considering these conflicts and the values of all \acp{ICP} presented in the table in Fig.~\ref{fig:example_model}, we generate a conflict table for each \ac{xApp} comprising all its associated \acp{ICP} and \acp{KPI} (see in Section~\ref{sec:kpi2uti} and the Github repository at \cite{datagithub}). The values of the \acp{KPI} for different control inputs of the parameters are estimated using the Gaussian distribution equation as it often mirrors real-life scenarios \cite{anubhab2020conflict, banerjee2021toward, wadud2023conflict}. Equations used for generating these KPIs are: $k_1 = 80 \times e^{-\frac{(p_1 + 0)^2}{2p_2^2}}$, $k_2 = 100 \times e^{-\frac{(p_1 + p_3)^2}{2p_2^2}}$, $k_3 = 120 \times e^{-\frac{(p_1 + 45)^2}{2p_4^2}}$, $k_{41} = 120 \times e^{-\frac{(p_6 + (p_2 -30))^2}{2p_5^2}}$, $k_{42} = 150 \times e^{-\frac{(p_6 + (p_2 -50))^2}{2p_5^2}}$, and $k_5 = -35 \times e^{-\frac{(p_8 + (p_1 -25))^2}{2p_7^2}}$. Afterwards, all \acp{xApp} are trained with an \ac{ANN} regression model to enhance each \ac{xApp}'s \ac{KPI} prediction capability for various settings of their \acp{ICP} using the generated dataset. Section~\ref{sec:kpi_pred} provides a detailed discussion on the prediction aspect. In the numerical analysis, QoS thresholds used for each of the KPIs are: $q_1 = 55,~ q_2 = 95,~ q_3 = 85,~ q_{41} = 75,~ q_{42} = 80,~ q_5 = -25$. 

\begin{table}[!ht]
 \centering
 \footnotesize
 \caption{List of Notations}
 \begin{tabular}{ p{1cm} p{7cm}}
 \hline
 \textbf{Symbol} & \textbf{Description} \\
 \hline
 \multicolumn{2}{l}{\textbf{Given Parameters:}} \\
 $X$ & Set of all \acp{xApp} in the \ac{Near-RT-RIC} \\
 $P$ & Set of \acp{ICP} associated with all \acp{xApp} in the \ac{Near-RT-RIC} \\
 $K$ & Set of all \acp{KPI} associated with all \acp{xApp} \\
 $k \in K$ & A particular \ac{KPI} belongs to an \ac{xApp} in the \ac{Near-RT-RIC} \\
 $Q$ & Set of \ac{QoS} thresholds for all \acp{xApp} \\
 $q \in Q$ & A particular \ac{QoS} threshold for an \ac{xApp} \\
 $Q^\prime$ & Set of \ac{QoS} thresholds for all conflicting \acp{xApp} \\
 $q^\prime \in Q ^\prime$ & A particular \ac{QoS} threshold for a conflicting \ac{xApp} \\
 $X^\prime$ & Set of all conflicting \acp{xApp} \\
 $x^\prime \in X^\prime$ & A particular conflicting \ac{xApp} \\
 $u$ & Utility of an \ac{xApp} converted from \acp{KPI} \\
 $U(p)$ & Utility function to convert \acp{KPI} to utility $u$ for a given value of the conflicting parameter $p$ using z-score normalization \\
 $w$ & Priority weight assigned by the \ac{CS} \ac{xApp} \\
 $d$ & Variable representing the shortfall of \ac{xApp}'s \ac{KPI} from its \ac{QoS} threshold \\
 $s$ & \ac{QoS} indicator of \acp{xApp} \\
 $\delta$ & Binary variable to indicate \acp{KPI} to be minimized or maximized \\
 $\zeta$ & Constant to balance the relative importance of the two parts of the objective function \\
 \hline
 \multicolumn{2}{l}{\textbf{Decision Variables:}} \\
 $p_l \in P$ & Value of the conflicting parameter \\
 \hline
 \end{tabular}
 \label{tab:notation}
\end{table}

\section{Conflict Management System Framework}
\label{sec:cms_framework}
In this article, we adopt the \ac{CMS} framework as presented in \cite{adamczyk2023conflict} and \cite{wadud2023conflict}. This concept, incorporating a database, shared data layer, and messaging infrastructure, is congruent with the existing \ac{Near-RT-RIC} architecture of the O-RAN ALLIANCE \cite{ric_oran_alliance}. Our recent work \cite{wadud2023conflict} proposed a conflict management system consisting of three primary components: the \ac{PMon}, the \ac{CDC}, and the \ac{CMC}. These components, together with the database, are integral to detecting and mitigating intra-component conflicts within the \ac{Near-RT-RIC}. The following sections discuss each of these components and the necessary database component, essential to the conflict management system framework.

\begin{figure}[!ht]
 \centering
 \vspace{-0.1in}
	\includegraphics[scale = 0.6]{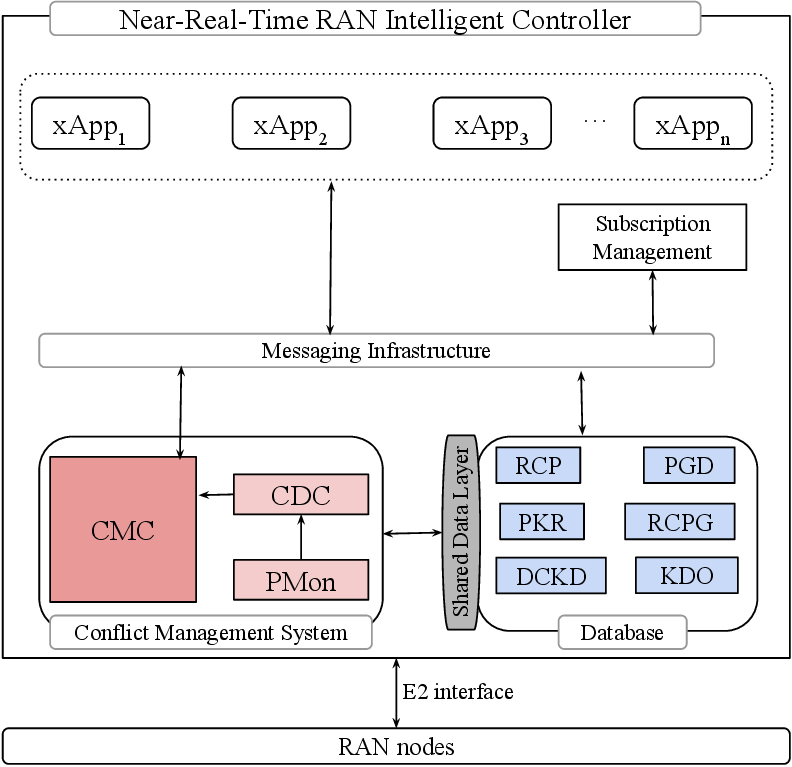}
	\caption{The CMS Framework in the \ac{Near-RT-RIC}.}
	\label{fig:cms_framework}
 \vspace{-0.1in}
\end{figure}
\subsubsection{Recently Changed Parameter (RCP)}
The \ac{RCP} component within the database archives all parameters recently modified at the behest of various \acp{xApp}. Each parameter is stored alongside its corresponding timestamp.

\subsubsection{Parameter Group Definition (PGD)}
Within this database segment, parameters impacting the same network zone are cataloged. For instance, parameters like antenna tilts and cell individual offset, which both influence a cell's handover boundary, are categorized together for each handover related \ac{KPI}.

\subsubsection{Recently Changed Parameter Group (RCPG)}
Changes to parameters within the \ac{PGD} are recorded in the \ac{RCPG} section of the database. Here, each parameter alteration is logged with its timestamp and the associated parameters within the same group.

\subsubsection{Parameter and KPI Ranges (PKR)}
The \ac{PKR} database section compiles the minimum and maximum permissible values for each parameter, along with the relevant \ac{KPI} for a specific cell.

\subsubsection{Decision Correlated with KPI Degradation (DCKD)}
This part of the database is dedicated to recording individual \ac{KPI} thresholds, which are determined based on the \ac{QoS} and \ac{SLA} requirements of the respective cells or networks.

\subsubsection{KPI Degradation Occurrences (KDO)}
The \ac{KDO} database component tracks occurrences of \ac{KPI} degradation. This tracking is done subsequent to modifications in parameters by the \acp{xApp} via \ac{RAN} nodes and includes respective timestamps for each change.

The following subsections detail the functionalities of the core \ac{CMS} components:

\subsubsection{Performance Monitoring Component (PMon)}\label{pmon}
The \ac{PMon} in the \ac{CMS} oversees monitoring and analysis of network \acp{KPI}. It gathers data from \ac{RAN} nodes through the E2 interface, including network elements and user devices (refer to Fig.~\ref{fig:cms_framework}). This data, encompassing measurements and statistics, aids in \ac{KPI} assessment against \ac{QoS} thresholds. Deviations in \ac{KPI} values, indicative of performance anomalies, are logged in the \ac{KDO} database. Upon \ac{KPI} breaches, \ac{PMon} alerts the \ac{CDC} to identify potential \acp{xApp} conflicts.

\subsubsection{Conflict Detection Controller (CDC)}\label{cdc}
The \ac{CDC}, as depicted in Fig.~\ref{fig:cms_framework}, detects various conflicts within the \ac{Near-RT-RIC}, including direct, indirect, and implicit types, essential for a robust Open \ac{RAN} architecture. Direct conflicts are identified through \acp{ICP} analysis during \acp{xApp} deployment, while indirect and implicit conflicts are recognized via \ac{KPI} degradation and parameter changes, respectively. The \ac{CDC} then informs the \ac{CMC} about any detected conflicts along with relevant details.

\subsubsection{Conflict Mitigation Controller (CMC)}\label{cmc}
The \ac{CMC} addresses intra-component conflicts in the \ac{Near-RT-RIC} by employing various conflict mitigation methods, including \ac{NSWF}, \ac{EG} or \ac{QACM}. In case of detected conflicts, it suggests new parameter values that maximize or minimize certain objective functions among the involved \acp{xApp}, based on their \acp{KPI}. The \ac{NSWF} and \ac{EG} discussed in Section~\ref{sec:bench} and \ac{QACM} in Section~\ref{sec:proposed} are used to calculate this optimal value, considering both the most recent and the previous \ac{KPI} values associated with the conflicting parameter.

In addition to these aforementioned components, we envision having a \ac{CS} \ac{xApp} deployed in the \ac{Near-RT-RIC} that provides the \ac{MNO} a solid control over the conflict mitigation system. The \ac{CS} \ac{xApp} closely monitors the network state and assigns weights to the conflicting \acp{xApp} upon requests from the \ac{CMC}. While assigning weights, it also considers the current policy configuration provided by the \ac{MNO}. The \ac{MNO} can update the policy configuration anytime that will immediately be effective to the \ac{CMS} control-loop. 

\subsection{Control Loop for the Conflict Management System}
\label{sec:cms_controlLoop}
\begin{figure}[!ht]
 \centering
 \vspace{-0.1in}
	\includegraphics[scale = 0.42]{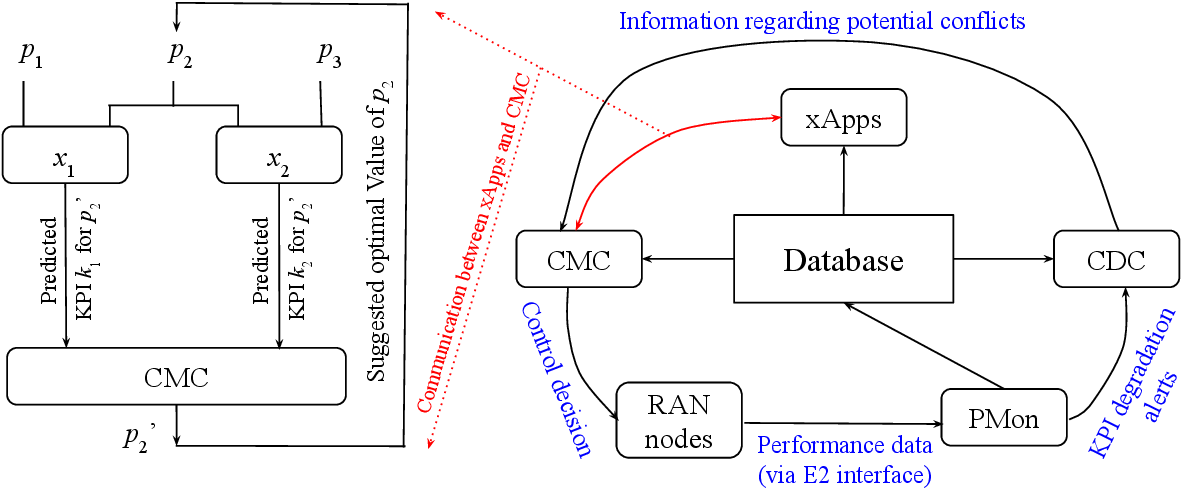}
	\caption{Control Loop for the \ac{CMS}.}
	\label{fig:cms_loop}
 \vspace{-20pt}
\end{figure}

The control loop of the \ac{CMS} and its interconnected database components is depicted in Fig.~\ref{fig:cms_loop}. This loop illustrates the systematic flow and interaction between different components within the \ac{CMS} highlighting the dynamic and responsive nature of the system. At the core of this loop is the constant monitoring and analysis of network performance, which is facilitated by the \ac{PMon} component. This component continuously gathers and processes data from various \ac{RAN} nodes, providing valuable insights into network performance through \ac{KPI} assessment.
Fig.~\ref{fig:cms_loop} further demonstrates the sequential flow of operations, starting from data collection to conflict resolution. Upon detecting deviations in \ac{KPI} values against predefined \ac{QoS} thresholds, the \ac{PMon} component triggers the \ac{CDC}. The \ac{CDC} then employs advanced algorithms to detect any potential conflicts between the \acp{xApp}. Following conflict identification, the \ac{CMC} takes charge, applying cooperative bargain game theory principles to resolve these conflicts, thereby ensuring optimal network performance. To facilitate this, the \ac{CMC} engages in an iterative communication process with the involved \acp{xApp} by exchanging information back and forth to calculate the optimal values of the conflicting parameters. This negotiation is visualized by the red line in the control loop depicted on the right side of Fig.~\ref{fig:cms_loop}. The \ac{CMC}--\acp{xApp} communication is illustrated on the left side of Fig.~\ref{fig:cms_loop}. The iterative bargaining process between the \ac{CMC} and the involved \acp{xApp} is crucial for reaching a consensus on the acceptable parameter values that satisfy the requirements of the network and the \acp{xApp} without compromising on the overall system performance. The entire process embodies a cohesive and efficient approach to maintaining network integrity and performance in the \ac{Near-RT-RIC}.
\section{Benchmark}
\label{sec:bench}
We consider \ac{NSWF} and \ac{EG} as the benchmark for performance comparison of the \ac{QACM} method. The former is used when all conflicting \acp{xApp} $x^\prime \in X^\prime$ have equal priority weights. The latter is used when different priority weights are assigned by the \ac{CS} \ac{xApp} based on the current network state and policy set by the \ac{MNO}. Nash's equilibrium is well-known for computing maximum collective utility among players or agents in a multi-player or multi-agent scenario. It simply estimates the product of utilities of individual agents involved in the game. The Eq.~\ref{eq:nswf} estimates the \ac{NSWF} for all conflicting \acp{xApp} $x_i^\prime \in X^\prime$ for the conflicting parameter $p_l$ by iterative bargain within its optimal configuration range (see  Section~\ref{sec:ocr}). 
\begin{equation}
NSWF(p_l) = \Pi_{i \in [1, |X^\prime|]} U_{i} (p_l), \forall i \in X^\prime \label{eq:nswf}
\end{equation}
The \ac{EG}, on the contrary, is used when particular \acp{xApp} require preferences over the other contentious \acp{xApp}. For instance, let us consider a practical scenario stated in Section~\ref{sec:application} where \ac{ES} and \ac{MRO} \ac{xApp} have a direct conflict over \ac{TXP}. If the current network state experiences high level of call drop rate, the \ac{CS} \ac{xApp} or the \ac{MNO} can decide to put more preference to \ac{MRO} over \ac{ES} and help to rise values for \ac{TXP} and \ac{CIO} to increase throughput and broaden the handover boundary. In such a case, \ac{EG} solution is essential, but \ac{NSWF} fails as it doesn't consider priority cases. The following is the \ac{EG} linear programming problem \cite{wadud2023conflict}: 

\begin{subequations}
\vspace{-15pt}
\begin{eqnarray}
& & \text{Maximize} \sum_{i \in [1, |X^\prime|]} w_i U_{i} (p_l) \label{eq:obj} \\
& {\rm s.t.} & \sum_{i \in [1, |X^\prime|]} w_{i}=1, \forall i \in X^\prime \label{eq:sum_weight} \\
& & p_l \geq p_l^{min, \rm{opt}}, \forall l \in P \label{eq:min_x} \\
& &  p_l \leq p_l^{max, \rm{opt}}, \forall l \in P  \label{eq:max_x} \\
& & \forall i \in [1, |X^\prime|], |X^\prime| \geq 2.  \label{eq:i_X}
\end{eqnarray}
\end{subequations}

The objective function \eqref{eq:obj} represents the goal of maximizing the weighted utility of individual \acp{xApp} within the set $X^\prime$ for a given parameter $p_l$. The utility function $U_i(p_l)$ captures the utility of \ac{xApp} $i$ for parameter $p_l$, which is influenced by specific network conditions and \ac{KPI} metrics. The constraint \eqref{eq:sum_weight} ensures that the weights $w_i$ assigned to each \ac{xApp} sum to 1, thereby normalizing the influence of each \ac{xApp} in the objective function. This normalization allows the aggregation of utilities to be proportionally representative of each \ac{xApp}'s importance. Constraints \eqref{eq:min_x} and \eqref{eq:max_x} define the permissible optimal range of the decision variable $p_l$. Finally, constraint \eqref{eq:i_X} ensures that the number of \acp{xApp} involved in the optimization is at least two. It highlights the focus on scenarios where multiple \acp{xApp} cooperate within the network.

\section{The Proposed Method}
\label{sec:proposed}
The proposed \ac{QACM} method is designed to be deployed as the \ac{CMC} component in the \ac{CMS}, as shown in Fig.~\ref{fig:cms_framework}. In this article, we assume that all other components of the \ac{CMS} framework perform their tasks efficiently, and the \ac{CMC} is notified by the \ac{CDC} when any direct, indirect, or implicit conflict occurs in the \ac{Near-RT-RIC}. Afterwards, the \ac{CMC} communicates back-and-forth with the conflicting \acp{xApp} $x^\prime \in X^\prime$ to ensure that the maximum number of \acp{xApp} meet their respective $q^\prime \in Q^\prime$ and to alleviate the negative impact of the transpired conflict. When an optimal value is obtained, achieving the objective goal of the proposed \ac{QACM} method while satisfying certain constraints (see in Section\ref{sec:qacm}), the \ac{CMC} forwards that optimal value of the conflicting \ac{ICP} as a control decision to the respective \acp{xApp} $x^\prime \in X^\prime$ and \ac{RAN} nodes. Certain prerequisites are necessary for the proposed \ac{QACM} method to be deployed in the \ac{Near-RT-RIC}, mainly the KPI prediction ability of individual \ac{xApp}. This ability can also be developed as an independent \ac{KPI} prediction \ac{xApp}. These prerequisites are discussed in Sections~\ref{sec:kpi2uti} and \ref{sec:kpi_pred}.

\subsection{KPI to Utility Conversion}
\label{sec:kpi2uti}
Converting \acp{KPI} to utilities and defining a common \ac{QoS} threshold for each \ac{xApp} in the presence of multiple \acp{KPI} associated with a single \ac{xApp} poses a significant challenge to the proposed \ac{QACM} method. We use z-score normalization \cite{colan2013and} technique for converting the associated \ac{KPI} of an \ac{xApp} to utility. The reason for using z-score normalization technique over the min-max normalization technique (used in \cite{wadud2023conflict}) is primarily because it maintains the Gaussian distribution of the original \ac{KPI} data, which is a crucial prerequisite for the proposed conflict mitigation model. Unlike min-max normalization, which simply re-scales the data to a fixed range, z-score retains the original distribution's properties, making it more suitable for utilities that rely on the underlying Gaussian-Normal characteristics. 

\begin{figure}[!ht]
 \centering
 \vspace{-0.1in}
	\includegraphics[scale = 0.4]{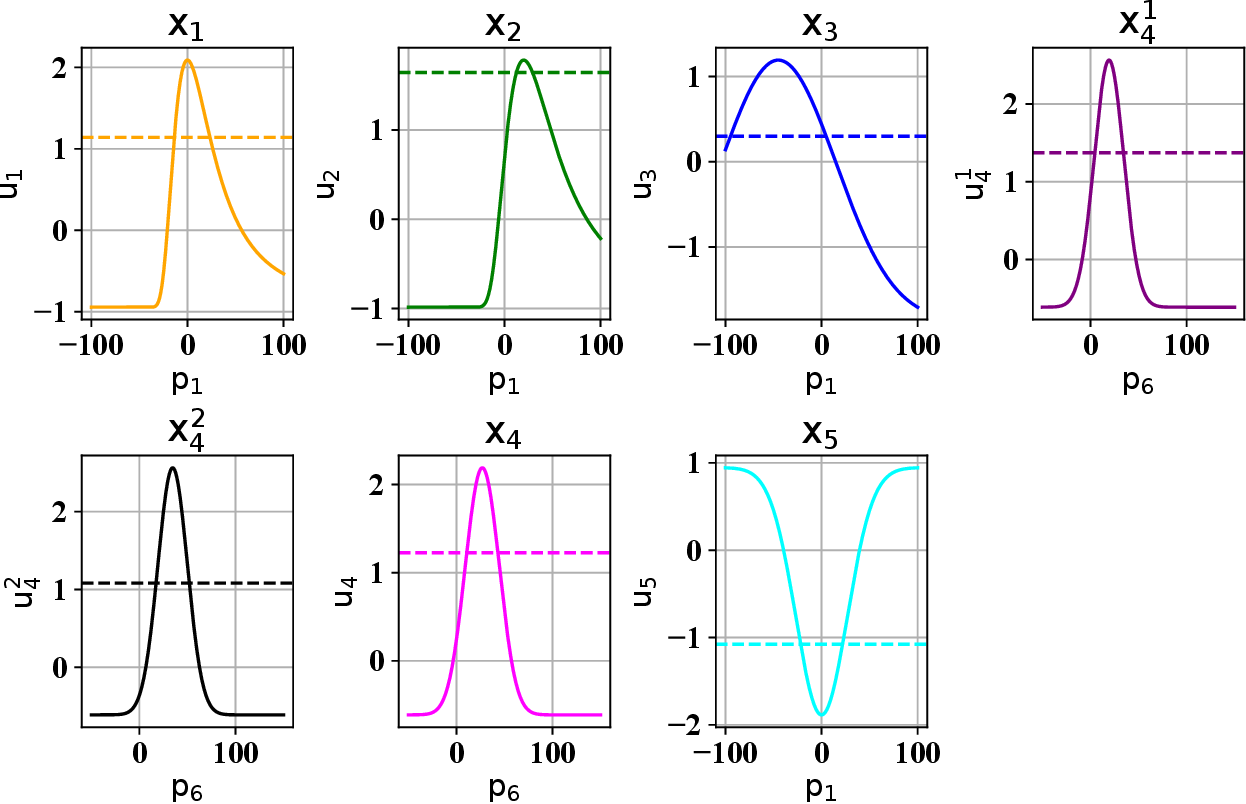}
	\caption{normalized \ac{KPI} values or utilities related to each \ac{xApp}.}
	\label{fig:kpi2utility}
 \vspace{-0.1in}
\end{figure}

To explain the conversion of \acp{KPI} to utility values, we consider an example model with five \acp{xApp} and their respective \ac{ICP} configurations as depicted in Fig.~\ref{fig:example_model}. The conflict tables for this model, including \acp{KPI} specific to each \ac{xApp} and their corresponding \ac{QoS} thresholds normalized using the z-score normalization technique, are available as datasets in our \textit{QACM repository} \cite{datagithub} \footnote{The dataset can be accessed directly via the following link: \href{https://github.com/dewanwadud1/QACM}{"QACM repository on GitHub"}. For guidance on how to interpret these conflict tables, please refer to the README file within the repository.}. Fig.~\ref{fig:kpi2utility} illustrates the normalized \acp{KPI} and their respective \ac{QoS} thresholds. All horizontal dashed lines in each subplot indicate the normalized QoS thresholds $q \in Q$, and continuous curved lines refer to \acp{KPI} $k \in K$.


In our example model in Fig.~\ref{fig:example_model}, each \ac{xApp} is linked to a single \ac{KPI} with the exception of \ac{xApp} $x_4$, which has multiple \acp{KPI}. This design choice reflects the complexity of real-world scenarios and demonstrates the necessity of converting \acp{KPI} into utility values. Because more number of \acp{KPI} associated with a single \ac{xApp} adds extra bargain weights to the mitigation method. In a real-life scenario, for example, common \acp{xApp} like \ac{CCO}, \ac{ES}, \ac{MRO}, and \ac{MLB} typically have multiple \acp{KPI}, as outlined in Tables~\ref{tab:kpi-xapps} and~\ref{tab:xapps-objectives}. Therefore, having a single \ac{KPI} representation for individual \ac{xApp} is necessary to keep the computational latency of the mitigation model within the \ac{Near-RT-RIC}'s latency budget. 
Normalization of \acp{KPI} into utilities results in a uniform scalar measure, usually ranging between $[-3, +3]$. In a Gaussian distribution, approximately 99.7\% of the data falls within three standard deviations from the mean. Therefore, after applying z-score normalization to the KPIs, which follow a Gaussian normal distribution, the resulting normalized scores predominantly range between -3 and +3. This standardisation allows for the integration of multiple \acp{KPI} associated with a single \ac{xApp} to a single measure through various statistical techniques. However, combining these \acp{KPI} requires careful consideration, as some may be more significant than others. Also, combining different \acp{KPI} together might not always be meaningful. Addressing the \ac{QoS} requirements of one \ac{KPI} can sometimes meet the needs of other \acp{KPI} related to the same \ac{xApp}. Thus, we emphasize the need for manual oversight in determining the relevant \acp{KPI} for conflict mitigation, given the complexity and contextual dependency of defining a \ac{KPI} importance hierarchy for each \ac{xApp}. This area, involving the identification of significant \acp{KPI} based on network conditions, presents substantial research opportunities, with potential applications for \ac{AI} and \ac{ML} methods which fall beyond the scope of the present work.
To represent all associated \acp{KPI} as a single vector, techniques like Principal Component Analysis (PCA) or Factor Analysis can also be adopted. However, in this paper, we utilize a weighted-average method, which averages the \acp{KPI} with weights assigned by the \ac{MNO}. The sixth figure from the top-left to the right in Fig.~\ref{fig:kpi2utility} presents the combined utility curve for the two \acp{KPI} of $x_4$, assuming equal importance for both. We presume that \acp{xApp} are pre-trained with offline \ac{KPI} prediction models capable of estimating \ac{KPI} values based on provided \acp{ICP}. However, we train \acp{xApp} with prediction models to evaluate the performance of the QACM framework and to demonstrate the prediction process, using the collected conflict table available at \cite{datagithub}, based on the example \ac{xApp} configuration shown in Fig.~\ref{fig:example_model}. This serves as a proof of concept for \ac{KPI} prediction that we plan to investigate further in our future work. The subsequent section delves into the prediction of \acp{KPI} for \acp{xApp} in greater detail.

\subsection{KPI Prediction for xApps}
\label{sec:kpi_pred}
\ac{KPI} prediction aids the \ac{CMC}--\acp{xApp} interaction process by helping to estimate the optimal value of the conflicting parameter (as discussed in Section~\ref{sec:cms_controlLoop} and illustrated in the left part of Fig.~\ref{fig:cms_loop}). We train two different regression models, \ac{ANN} and \ac{PR}, for each \ac{xApp} with individual conflict tables collected from the example model shown in Fig.~\ref{fig:example_model}. We use simplified models for \ac{KPI} prediction as a proof of concept for the underlying principles of the \ac{QACM} framework. More sophisticated models are required for \ac{KPI} prediction in actual network environment \cite{tran2023ml}. We discussed more about this in Section~\ref{sec:limit_fut}. The \ac{ANN} comprises four hidden layers, each with 128 neurons. These layers utilize the hyperbolic tangent ($tanh$) activation function which is known for its efficacy in capturing non-linear relationships in the data. To mitigate the risk of overfitting, a dropout layer with a dropout rate of $0.2$ is included after each hidden layer. This setup randomly sets a fraction of input units to $0$ during training, helping to prevent complex co-adaptations on training data. The model terminates in an output layer with a single neuron employing a $linear$ activation function that is ideal for regression tasks. For compiling the model, we used the $adam$ optimiser that is a popular choice for deep learning applications, and the mean squared error loss function that is standard for regression problems. The model was trained over 10 epochs with a batch size of 10, balancing the efficiency and learning capability. 


\begin{table}[h]
\centering
\caption{Performance Comparison between ANN and PR}
\label{tab:regression_performance}
\begin{tabular}{|c|ccc|ccc|}
\hline
\multirow{2}{*}{xApps} & \multicolumn{3}{c|}{PR} & \multicolumn{3}{c|}{ANN} \\ \cline{2-7} 
    & EVS       & R-squared   & MSE       & EVS       & R-squared   & MSE       \\ \hline
$x_1$ & 0.82 & 0.82 & 0.18 & 0.96 & 0.95 & 0.04 \\
$x_2$  & 0.93 & 0.93 & 0.067 & 0.99 & 0.99 & 0.0006 \\
$x_3$  &  0.99  & 0.99 & 0.0003 & 0.99 & 0.99 & 0.0045 \\
$x_4$  &  0.96  & 0.96 & 0.02 & 0.98 & 0.98 &  0.01 \\
$x_5$ &  0.81  & 0.81 & 0.185 & 0.98 & 0.98 & 0.015 \\ \hline
\end{tabular}
\vspace{-10pt}
\end{table}
We compared the performance of \ac{ANN} and \ac{PR} models across five different \acp{xApp}. The results as summarized in Table~\ref{tab:regression_performance} indicate that the ANN model generally outperformed the PR model in terms of \ac{EVS}, R-squared, and \ac{MSE}. Specifically, the ANN model demonstrated higher \ac{EVS} and R-squared values, suggesting it was more effective in capturing the variance and predicting the outcomes accurately for most \acp{xApp}. Moreover, the ANN model exhibited lower \ac{MSE} values, indicating that its predictions were closer to the actual values compared to those of the \ac{PR} model. Therefore, we decide to use the \ac{ANN} model for predicting \acp{KPI} of individual \ac{xApp}. 
\subsection{Estimating the Optimal Configuration Range}
\label{sec:ocr}
When the \ac{CMC} is informed about a conflict by the \ac{CDC}, it immediately communicates with the set of involved \acp{xApp} in $X^\prime$ and the \ac{CS} \ac{xApp}. The \ac{CS} \ac{xApp} quickly assesses the network policy configuration set by the \ac{MNO} and the current network state, and subsequently, provides weights $w_i$ for each involved \ac{xApp} $x_i^\prime \in X^\prime$ so that $\sum_{i \in [1, |X^\prime|]} w_{i}=1$. Simultaneously, all the involved \acp{xApp} $x^\prime \in X^\prime$ are asked to provide the individual optimal configuration range of the conflicting parameter $p_l$ as $\{p_l^{min, \rm{x}_j^\prime} ,p_l^{max, \rm{x}_j^\prime}\}$, where $p_l \in P$ and $x_j^\prime \in X^\prime$. Afterwards, the \ac{CMC} estimates the overall optimal configuration range $\{p_l^{min, opt}, p_l^{max, opt}\}$ of the conflicting parameter $p_l$. Suppose, there are two \acp{xApp} $x_1^\prime$ and $x_2^\prime$ involved in a particular conflict over parameter $p_1$. Upon request from the \ac{CMC}, $x_1^\prime$ and $x_2^\prime$ send their individual optimal configuration range $\{p_1^{min, \rm{x}_1^\prime}, p_1^{max, \rm{x}_1^\prime}\}$ and $\{p_1^{min, \rm{x}_2^\prime}, p_1^{max, \rm{x}_2^\prime}\}$. The \ac{CMC} estimates the optimal configuration range as $\{p_l^{min, opt}, p_l^{max, opt}\} \simeq \{min(p_1^{min, \rm{x}_1^\prime}, p_1^{min, \rm{x}_2^\prime}), max(p_1^{max, \rm{x}_1^\prime}, p_1^{max, \rm{x}_2^\prime})\}$.

\subsection{QACM Method}
\label{sec:qacm}
\begin{figure}[!ht]
 \centering
 \vspace{-0.1in}
	\includegraphics[scale = 0.37]{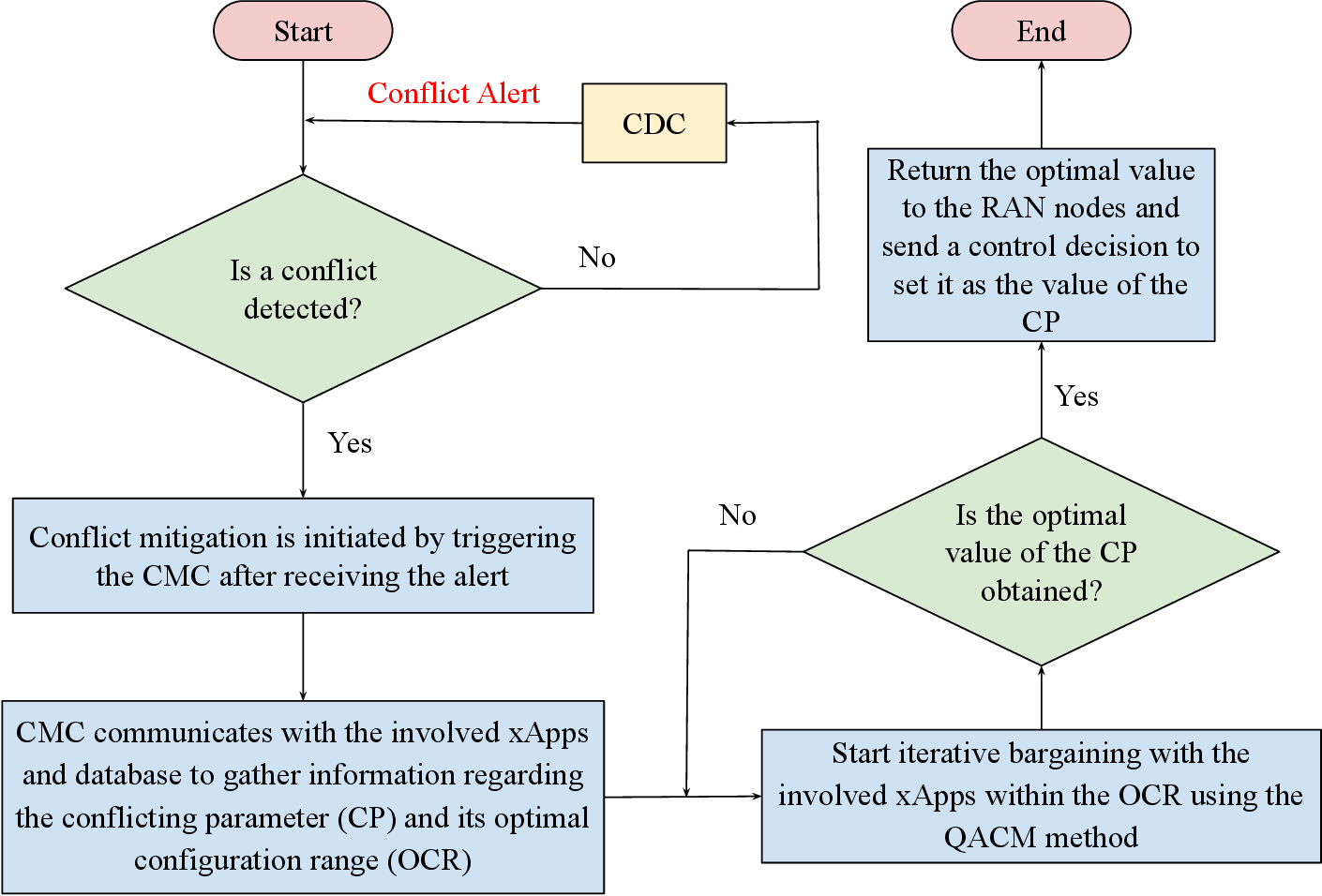}
	\caption{Flow diagram of conflict mitigation with the proposed \ac{QACM} method.}
	\label{fig:qacm_flow}
 \vspace{-0.1in}
\end{figure}

The goal of the proposed \ac{QACM} method is to minimize weighted distances of \acp{xApp} \acp{KPI} from their respective \ac{QoS} thresholds and a squared sum of \ac{QoS} satisfaction indicators. This method estimates the optimal value of the conflicting parameter, which is later passed as a control decision to underlying \ac{RAN} nodes by the \ac{CMC}. Fig.~\ref{fig:qacm_flow} illustrates the flow diagram of conflict mitigation for the proposed method. We formulate the following optimization problem for \ac{QACM} in this regard:
\allowdisplaybreaks
\begin{subequations}
\begin{align}
    & \text{Minimize} \quad \sum_{i \in [1, |X^\prime|]} w_i d_i \times \zeta - (\sum_{i \in [1, |X^\prime|]} s_i)^2 \label{eq:obj_qacm} \\
    & \text{s.t.} \quad \sum_{i \in [1, |X^\prime|]} w_{i} = 1, \label{eq:qacm_sum_weight} \\
    & \sum_{i \in [1, |X^\prime|]} s_i \leq |X^\prime|, \label{eq:qos_sum} \\
    & d_i \geq 0 \quad \forall i \in [1, |X^\prime|], \label{eq:si_non_negative} \\
    & d_i \geq (q_i^\prime - U_i(p_l)) \times (1 - \delta_i) + (U_i(p_l) - q_i^\prime) \times \delta_i \quad \forall i \in [1, |X^\prime|], \label{eq:si_slack} \\
    & U_i(p_l) \geq q_i^\prime - M(1 - s_i) \quad \forall i \in [1, |X^\prime|], \label{eq:ui_threshold} \\
    & s_i \in \{0, 1\} \quad \forall i \in [1, |X^\prime|], \label{eq:zi_binary} \\
    & \delta_i \in \{0, 1\} \quad \forall i \in [1, |X^\prime|], \label{eq:delta_binary} \\
    & p_l^{min, \rm{opt}} \leq p_l \leq p_l^{max, \rm{opt}}, \quad \forall p_l \in P, \label{eq:pl_bounds} \\
    & |X^\prime| \geq 2. \label{eq:minConf}
\end{align}
\end{subequations}

The Eq.~\eqref{eq:obj_qacm} represents the objective function of the optimization problem. It minimizes the weighted sum of distance reduced by the squared sum of satisfaction indicators $s_i$. It balances the need to keep \acp{xApp} within desired \ac{QoS} levels while maximizing overall satisfaction. The Eq.~\eqref{eq:qacm_sum_weight} is the constraint that ensures the total sum of weights $w_i$ assigned to each \ac{xApp} is equal to 1. The total satisfaction score is constrained to the number of \acp{xApp} by the Eq.~\eqref{eq:qos_sum}. Eq.~\eqref{eq:si_non_negative} and Eq.~\eqref{eq:si_slack} ensure that \( d_i \) is non-negative and measures the deviation from \( q_i^\prime \). Specifically, \( d_i \) measures the shortfall when \(\delta_i = 0\) (KPI maximized) and the excess when \(\delta_i = 1\) (KPI minimized). Eq.~\eqref{eq:delta_binary} ensures \(\delta_i\) is binary, with \(\delta_i = 1\) for KPIs to be minimized and \(\delta_i = 0\) for KPIs to be maximized. Binary condition in Eq.~\eqref{eq:ui_threshold} and Eq.~\eqref{eq:zi_binary} ensure \( s_i \) accurately reflects compliance with QoS thresholds using a large constant \( M \) to enforce the binary condition. $M$ is larger than the maximum possible value of \( U_i(p_l) \). A bound is applied on the decision variable $p_l$ in Eq.~\eqref{eq:pl_bounds}. The permissible range for the decision variable is within the optimal configuration range discussed in Section~\ref{sec:ocr}. The final constraint in Eq.~\eqref{eq:minConf} indicates that the number of \acp{xApp} involved in this conflict mitigation approach is at least two because we consider scenarios involving at least two different \acp{xApp} in a conflicting setting.

\subsubsection{Complexity Analysis}
\label{sec:qacm_complexity}
In our optimization problem, the decision variable is primarily $p_l$, giving us only a unit variable. Additionally, we have $w_i$, $s_i$, and$\delta_i$ for each $i \in [1, |X^\prime|]$, contributing twice to $|X^\prime|$ variables. The $d_i$ for each $i$ also adds twice to $|X^\prime|$ variables, leading to a total of $4|X^\prime|$ variables.


The number of constraint analyses is equally straightforward. We have a single weighted sum constraint in Eq.~\eqref{eq:qacm_sum_weight}, a single satisfaction sum constraint in Eq.~\eqref{eq:qos_sum}, and $|X^\prime|$ constraints each from the distance conditions in Eq.~\eqref{eq:si_non_negative} and Eq.~\eqref{eq:si_slack}. Similarly, $|X^\prime|$ constraints each from the satisfaction indicator condition in Eq.~\eqref{eq:ui_threshold}, Eq.~\eqref{eq:zi_binary} and Eq.~\eqref{eq:delta_binary}. Adding $2|N|$ constraints for the bounds on decision variables in Eq.~\eqref{eq:pl_bounds} considering $|N|$ discrete values for $p_l$ within the range $[p_l^{min, \text{opt}}, p_l^{max, \text{opt}}]$ and one for the minimum \acp{xApp} involved in conflict in Eq.~\eqref{eq:minConf}, we reach a total of \( 4|X^\prime| + 2|N| + 3 \) constraints. This analysis highlights the computational complexity inherent to this latency-sensitive system.

\subsection{QACM in a Dynamic Environment}
\label{sec:dynamic_qacm}
For a dynamic scenario in the \ac{CMC}, where the optimization problem is not tractable for large instance of conflicting \acp{xApp} and optimal configuration bound of the conflicting parameter $p_l$, a heuristic approach can be adopted. By not tractable for large instances, we specifically refer to scenarios involving a large number of conflicting \acp{xApp} and a wide range of optimal parameter configurations for $p_l$, which significantly increase the computational complexity. The complexity analysis in Section~\ref{sec:qacm_complexity} shows that exponential behavior can manifest as the size of $|X^\prime|$ and $|N|$ increases. Particularly, when each \ac{xApp} interacts with the \ac{CMC} during iterative bargaining, the number of interactions and the complexity of calculating the optimal value of $p_l$ can grow exponentially. We developed Alg.~\ref{algo:alg_qacm} in light of the \ac{QACM} optimization problem as follows:


\begin{algorithm}[!ht]
\caption{Heuristic \ac{QACM} for dynamic environment}
\begin{algorithmic}[1]
\REQUIRE $X^\prime$, bound \{$p_l^{min, \text{opt}}, p_l^{max, \text{opt}}$\} for $p_l$, $w_i$ and $q_i^\prime$ for $\forall i \in [1, |X^\prime|]$, $\zeta$, $\delta$
\ENSURE Optimal value for $p_l$, $p_l^{\text{opt}}$.
\STATE Initialize a $cost$ array.
\STATE Initialize a $p_l^{\text{opt}}$ variable.
\STATE Initialize $fCost = 0$, $minCost = \infty$ variables.
\STATE Initialize an array $s$ as \ac{QoS} indicator.
\FOR{each $p_l \in [p_l^{min, \text{opt}}, p_l^{max, \text{opt}}]$}
\FOR{each $i \in [1, |X^\prime|]$}
\STATE Obtain predicted $U_i(p_l)$ from $x_i^\prime \in X^\prime$
\IF{$\delta_i = 0$}
    \IF{$U_i(p_l) < q_i^\prime$}
    \STATE Update $d_i = q_i^\prime - U_i(p_l)$
    \STATE Update $s[i] = 0$
    \ELSE
    \STATE Update $d_i = 0$
    \STATE Update $s[i] = 1$
    \ENDIF
\ELSE
    \IF{$U_i(p_l) > q_i^\prime$}
    \STATE Update $d_i = U_i(p_l) - q_i^\prime$
    \STATE Update $s[i] = 0$
    \ELSE
    \STATE Update $d_i = 0$
    \STATE Update $s[i] = 1$
    \ENDIF
\ENDIF
\STATE $cost[i] = w_i d_i \times \zeta$
\ENDFOR
\STATE $fCost =  \sum(cost) - (\sum(s))^2$
\IF{$minCost > fCost$}
\STATE Update $minCost = fCost$
\STATE Update $p_l^{\text{opt}} = p_l$.
\ENDIF
\ENDFOR
\RETURN $p_l^{\text{opt}}$.
\end{algorithmic}
\label{algo:alg_qacm}
\end{algorithm}

In Alg.~\ref{algo:alg_qacm}, the set of conflicting \acp{xApp} $X^\prime$, the optimal bounds \{$p_l^{min, \text{opt}}, p_l^{max, \text{opt}}$\} for the conflicting parameter $p_l$, weights assigned by the \ac{CS} \ac{xApp} $w_i$, and the $QoS$ threshold $q_i^\prime$ for $\forall i \in [1, |X^\prime|]$ are required as inputs. From steps~1 to 4, Alg.~\ref{algo:alg_qacm} initializes the required variables for estimating the $minCost$ and the optimal value for $p_l$. The outer loop from steps~5 to 26 iterates within the range \{$p_l^{min, \text{opt}}, p_l^{max, \text{opt}}$\}, and the inner loop from steps~6 to 21 iterates for each conflicting \ac{xApp} $x_i^\prime \in X^\prime$. The inner loop obtains $U_i(p_l)$ for an $x_i^\prime \in X^\prime$ in step~7, calculates the distance between $q_i^\prime$ and $U_i(p_l)$ based on the value of $\delta_i$. If $\delta_i = 0$, the distance is measured as $q_i^\prime - U_i(p_l)$ when $U_i(p_l) < q_i^\prime$, otherwise it is set to 0. If $\delta_i = 1$, the distance is measured as $U_i(p_l) - q_i^\prime$ when $U_i(p_l) > q_i^\prime$, otherwise it is set to 0. The \ac{QoS} indicator $s_i$ is updated accordingly from steps~8 to 21, and estimates the weighted distance and stores it in the $cost$ array in step~22. The algorithm breaks out from the inner loop in step~22. The final cost $fCost$ of all \acp{xApp} for a particular value of $p_l$ within the range \{$p_l^{min, \text{opt}}, p_l^{max, \text{opt}}$\} is calculated in step~23. If the calculated $fCost$ is greater than $minCost$, the value of $minCost$ and $p_l^{\text{opt}}$ is updated using steps~24 to 26. When all iterations of the outer loop are completed, Alg.~\ref{algo:alg_qacm} returns the final $p_l^{\text{opt}}$ in step~27.

\subsubsection{Complexity Analysis}
The complexity of Alg.~\ref{algo:alg_qacm} is primarily determined by its two nested loops. The outer loop iterates over each potential value of $p_l$ within the range $[p_l^{min, \text{opt}}, p_l^{max, \text{opt}}]$. The inner loop, executed within each iteration of the outer loop, runs for each conflicting \ac{xApp} in $X^\prime$, giving it a complexity of $O(|X^\prime|)$. Each iteration involves a series of calculations, including obtaining predicted utilities, updating distances, and modifying \ac{QoS} indicators and costs. Assuming the range of $p_l$ consists of $N$ discrete values, the overall complexity of the algorithm is approximately $O(N \cdot |X^\prime|)$.

\section{Case Study}
\label{sec:case_study}
We consider four conflicting cases for analyzing the performance of the proposed \ac{QACM} method as shown in Fig.~\ref{fig:example_model}. We use \ac{QACMP} in the resulting figures only for comparing priority cases with the result of \ac{EG} method. It is the same \ac{QACM} method with different priority weights for involved \acp{xApp}. In this section, the \ac{QACM} illustrates results for a conflicting scenario where each of the conflicting \ac{xApp} has equal weight, and it is compared with the result of \ac{NSWF} method. All numerical analysis were conducted on a Python-based simulator same as \cite{wadud2023conflict}. The following critically analyzes the performance of the proposed model and compares its performance with two specific benchmarks as stated above.

\subsection{Direct conflict between two xApps}
\label{sec:direct_two}
\begin{figure}[!ht]
 \centering
 \vspace{-0.1in}
	\includegraphics[scale = 0.4]{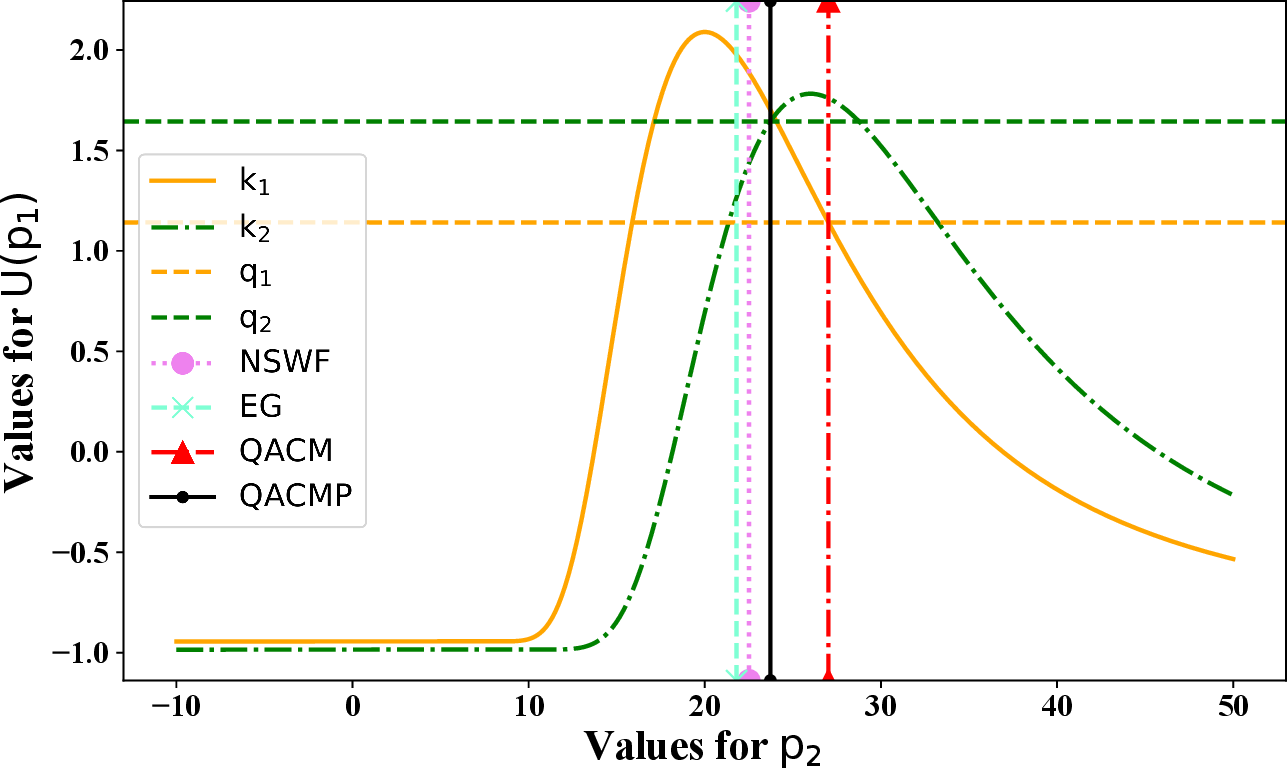}
	\caption{Direct conflict between $x_1$ and $x_2$ over $p_2$.}
	\label{fig:dConP2}
 \vspace{-0.1in}
\end{figure}

Fig.~\ref{fig:dConP2} illustrates a conflicting case between $x_1$ and $x_2$ over $p_2$. The curved lines, $k_1$ and $k_2$, indicate the utilities belonging to each \ac{xApp}, and the dashed horizontal straight lines, $q_1$ and $q_2$, represent their respective \ac{QoS} thresholds. The utility $k_1$ peaks at $p2 = 18$, leading $x_1$ to request the \ac{RIC} to set $p_2 = 18$. Conversely, $x_2$ seeks to set $p_2 = 25$ to reach its maximum utility. To resolve this direct conflict, we execute the \ac{QACM}, \ac{QACMP}, \ac{NSWF}, and \ac{EG} methods. In a non-priority scenario with equal priority weights, $w_1 = 0.5$ and $w_2 = 0.5$, the \ac{QACM} suggests setting $p_2 \approx 27$, where both \acp{xApp} meet their individual \ac{QoS} thresholds $q_1$ and $q_2$, respectively. In contrast, the \ac{NSWF} suggests setting $p_2 \approx 23$, where only $x_1$ meets its requirement $q_1$, but $x_2$ fails to meet $q_2$. In a priority scenario, with $w_1 = 0.7$ and $w_2 = 0.3$, the \ac{QACMP} suggests setting $p_2 \approx 25$, where both involved \acp{xApp} meet their \ac{QoS} thresholds. Conversely, the \ac{EG} method, under the same priority configuration, suggests setting $p_2 \approx 22$, where only $x_1$ meets $q_1$. In both scenarios, the \ac{QACM} ensures that the maximum number of involved \acp{xApp} meet their individual \ac{QoS} requirements.


\subsection{Direct conflict among multiple xApps}
\label{sec:direct_multiple}
\begin{figure}[!ht]
 \centering
 \vspace{-0.1in}
	\includegraphics[scale = 0.4]{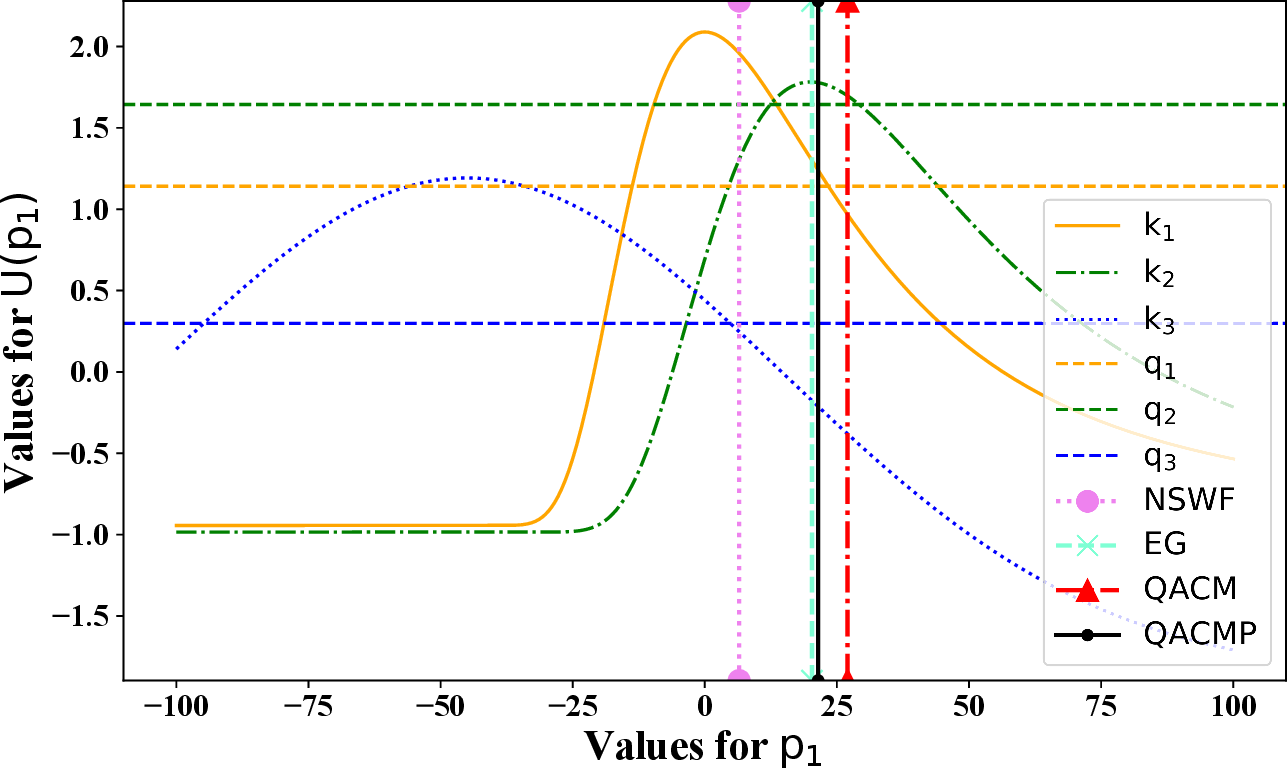}
	\caption{Direct conflict among $x_1$, $x_2$ and $x_3$ over $p_1$.}
	\label{fig:dConP1}
 \vspace{-0.1in}
\end{figure}

As depicted in Fig.~\ref{fig:dConP1}, a conflict scenario involving three \acp{xApp} over parameter $p_1$ is presented. In this case, the utility curves of the \acp{xApp}, denoted as $k_1$, $k_2$, and $k_3$, are shown along with their respective \ac{QoS} thresholds, represented by dashed horizontal lines $q_1$, $q_2$, and $q_3$. The utility curve $k_1$ for \ac{xApp} $x_1$ reaches its maximum at a certain value of $p_1 = 0$, prompting $x_1$ to request the \ac{RIC} to set $p_1 = 0$ to this optimal point. Similarly, \acp{xApp} $x_2$ and $x_3$ each prefer different settings of $p_1$, as $p_1 = 20$ and $p_1 = -45$, respectively, to maximize their respective utilities. This scenario leads to a complex conflict among the three \acp{xApp}, each vying for a different configuration of $p_1$.

To address this tripartite conflict, we apply the \ac{QACM} method alongside the \ac{NSWF} and \ac{EG} solutions for both priority and non-priority settings. In a non-priority setting with equal weights for all \acp{xApp}, the \ac{QACM} method effectively finds a configuration of $p_1 \approx 23$ that satisfies the \ac{QoS} requirements of $x_1$ and $x_2$. The \ac{NSWF} methods, under similar conditions, suggest different configurations of $p_1 \approx -45$, favoring only $x_3$. In contrast, in a priority setting, the \ac{QACM} method adapts to the assigned weights $\{w_1 = 0.1, w_2 = 0.2, w_3 = 0.7\}$ and finds an optimal configuration of $p_1 \approx 5$ that prioritizes and meets the \ac{QoS} requirements of $x_3$ while meeting the same for $x_1$. At the same time, it keeps the deviation of $k_2$ from $q_2$ as smaller as possible. The \ac{EG} method for the same setting finds an optimal configuration of $p_1 \approx 1$ that meets \ac{QoS} requirements of both $x_1$ and $x_3$, but increases deviation between $k_2$ and $q_2$. This figure and its analysis underscore the efficacy of the \ac{QACM} method in resolving conflicts involving multiple \acp{xApp} with varying \ac{QoS} requirements. It demonstrates the method's capability to navigate complex multi-party conflicts and identify configurations that balance the competing needs of different \acp{xApp} in the \ac{Near-RT-RIC}.

\subsection{Concurrent Mitigation of  direct and indirect conflicts}
\label{sec:direct_indirect}
\begin{figure}[!ht]
 \centering
 \vspace{-0.1in}
	\includegraphics[scale = 0.4]{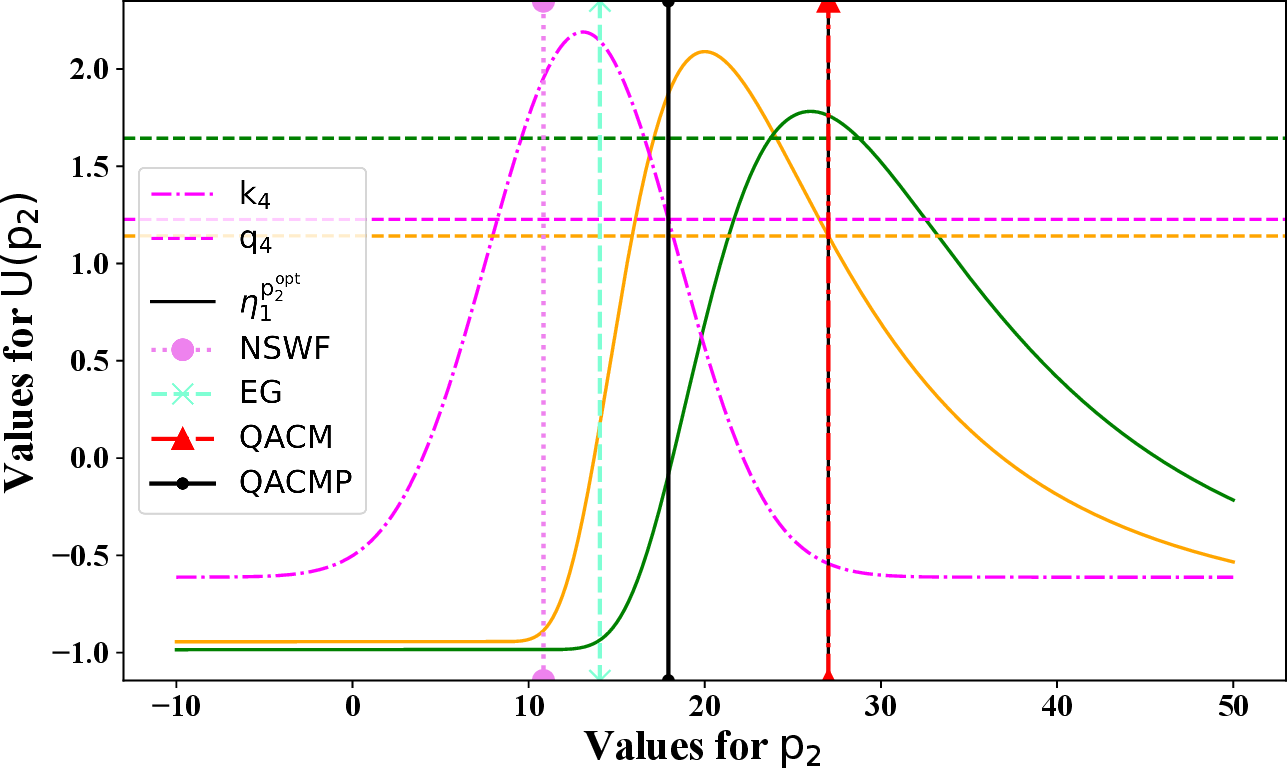}
	\caption{Indirect conflict of $x_4$ with $x_1$ and $x_2$ over $p_2$.}
	\label{fig:IndConP2}
 \vspace{-0.1in}
\end{figure}

Fig.~\ref{fig:IndConP2} illustrates two distinct network states, $\eta_1$ and $\eta_2$. The former, $\eta_1$, refers to the network situation when a direct conflict over $p_2$ occurred between $x_1$ and $x_2$. The latter, $\eta_2$, indicates a later situation when the suggested value of $p_2$ by the \ac{QACM} method to resolve the former conflict induces an indirect conflict with $x_4$. Within the context of Fig.~\ref{fig:example_model}, $p_2$ is implicated in a direct conflict between $x_1$ and $x_2$, as well as an indirect conflict involving $x_4$. In network state $\eta_1$, the \ac{QACM} non-priority method suggests a solution, $\eta_1^{p_2^{opt}} \approx 27$, to resolve the direct conflict, as depicted in Fig.~\ref{fig:dConP2}. However, this solution inadvertently diminishes the utility of $x_4$ below its \ac{QoS} threshold $q_4$, highlighting the indirect impact of $p_2$ on the \ac{KPI}s of $x_4$. Consequently, $x_4$ advocates for setting $p_2$ to 15, where it achieves maximum utility, introducing a new conflict with $x_1$ and $x_2$.

The \ac{QACM} method, along with the \ac{NSWF} and \ac{EG} solutions, is applied in both priority and non-priority settings. In a non-priority setting, the \ac{QACM} and \ac{NSWF} suggest $p_2 \approx 27$ and $p_2 \approx 11$, respectively, represented as $\eta_2^{QACM}$ and $\eta_2^{NSWF}$ in Fig.~\ref{fig:IndConP2}. The $\eta_2^{QACM}$ maintains the previously suggested value, ensuring both $x_1$ and $x_2$ meet their respective \ac{QoS} thresholds, due to the \ac{QACM} method's inherent mechanism of maximizing the number of \acp{xApp} that meet their individual \ac{QoS} requirements.

Conversely, in a priority setting, the \ac{QACM} method adjusts to the weights $\{w_1 = 0.1, w_2 = 0.2, w_4 = 0.7\}$ and identifies an optimal $p_2 \approx 18$ that satisfies $q_1$ and $q_4$, but not $q_2$. The \ac{EG} method, with the same weights, proposes $p_2 \approx 14$, meeting only $q_4$. Thus, the \ac{QACM} method surpasses both the \ac{NSWF} and \ac{EG} in non-priority and priority scenarios, respectively. This example showcases the intricate dynamics of conflict in Open \ac{RAN} and emphasizes the \ac{QACM} method's capacity to reconcile diverse \ac{QoS} requirements among multiple \acp{xApp}.

\subsection{Concurrent Mitigation of direct and implicit conflicts}
\label{sec:direct_implicit}

\begin{figure}[!ht]
 \centering
 \vspace{-0.1in}
	\includegraphics[scale = 0.4]{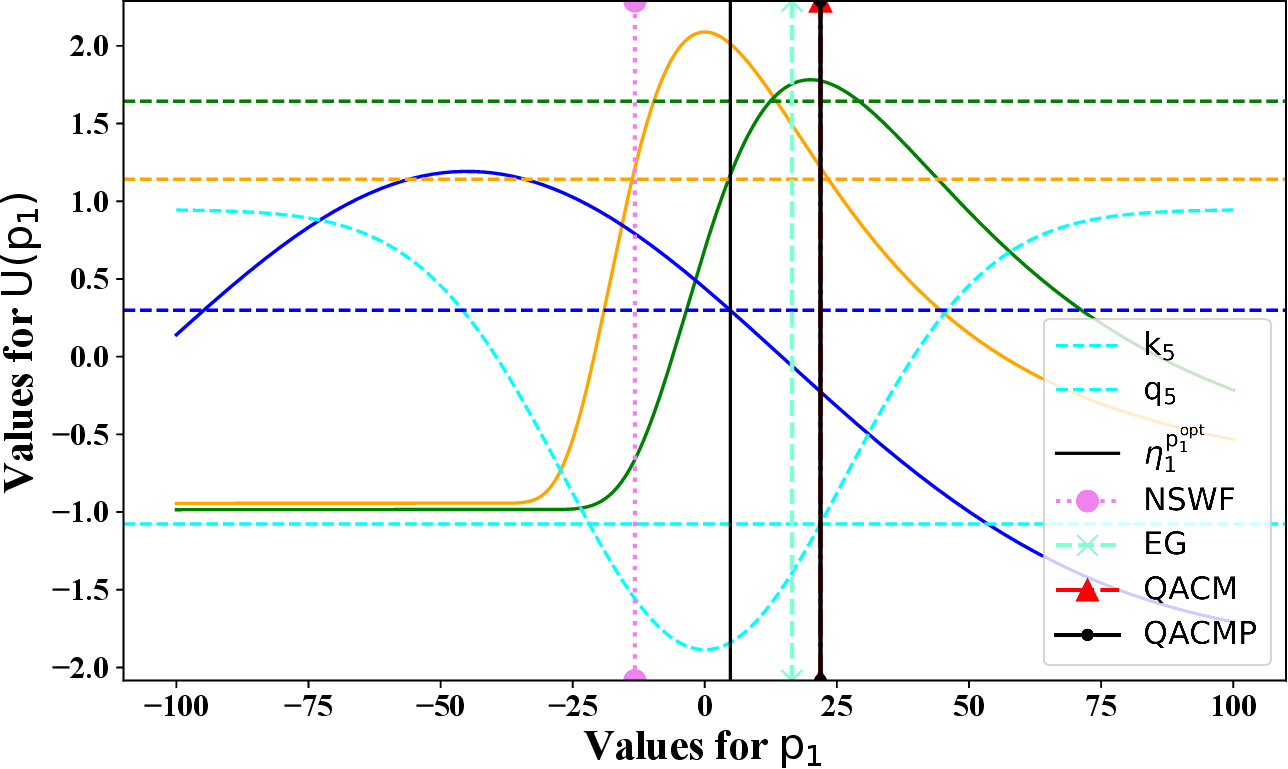}
	\caption{Implicit Conflict of $x_5$ with $x_1$, $x_2$ and $x_3$ over $p_1$.}
	\label{fig:ImpConP1}
 \vspace{-0.1in}
\end{figure}

Fig.~\ref{fig:ImpConP1} illustrates a scenario in which an implicit conflict, coupled with a direct conflict, emerges over the parameter $p_1$. We consider two network states, $\eta_1$ and $\eta_2$, akin to the previous case. Network state $\eta_1$ corresponds to the situation depicted in Fig.~\ref{fig:dConP1}, where the \ac{QACMP} scheme yields $p_1 \approx 5$ as the optimal solution for the conflict involving $x_1$, $x_2$, and $x_3$ over $p_1$. However, this resolution leads to a new conflict due to its detrimental effect on the utility of $x_5$. This conflict is presumed to be identified by the \ac{CDC}, prompting the execution of the \ac{QACM}, \ac{EG}, and \ac{NSWF} methods to determine an optimal value for $p_1$ once more, this time considering all four \acp{xApp}. This is referred to as network state $\eta_2$.

In network state $\eta_2$, the \ac{QACM} method and the \ac{NSWF} provide solutions for $p_1$ in the non-priority case, with $\eta_2^{QACM} \approx 22$ and $\eta_2^{NSWF} \approx -13$, respectively. The \ac{QACM} satisfies the \ac{QoS} requirements for three \acp{xApp}: $x_1$, $x_2$, and $x_5$. In contrast, \ac{NSWF} fulfills the requirements for two \acp{xApp}: $x_1$ and $x_3$. In the priority case, with the configuration $\{w_1 = 0.1, w_2 = 0.2, w_3 = 0.3, w_5 = 0.4\}$, the \ac{QACMP} suggests the same value as \ac{QACM}, while \ac{EG} calculates $p_1 \approx 16$. Similar to the non-priority case, \ac{QACMP} outperforms \ac{EG} in meeting the \ac{QoS} requirements of the involved \acp{xApp}. This scenario underscores the complexities involved in simultaneously addressing direct and implicit conflicts.


\section{Practical Application}
\label{sec:application}
This article primarily focuses on the theoretical development of a \ac{QoS}-aware conflict mitigation approach. To illustrate the application of the proposed method in a real-life scenario, we consider four \acp{xApp}, including \ac{CCO}, \ac{ES}, \ac{MRO}, and \ac{MLB}, as presented in Table~\ref{tab:kpi-xapps} and their respective objectives in Table~\ref{tab:xapps-objectives}. From the list of \acp{ICP} in Table~\ref{tab:xapps-objectives}, we identify four direct conflicts. Firstly, all four \acp{xApp} share the \ac{TXP}, leading to a direct conflict over \ac{TXP}. Secondly, \ac{MLB} and \ac{CCO} have a direct conflict over \ac{RET}. Lastly, \ac{MLB} and \ac{MRO} have two direct conflicts over \ac{CIO} and \ac{TTT}, respectively. Additionally, there is an indirect conflict between \ac{MLB} and \ac{MRO} \acp{xApp}. All \acp{KPI} related to handover are influenced by this group of parameters: \{\ac{TTT}, \ac{CIO}, \ac{TXP}, \ac{RET}, \ac{HYS}\}. Any change in these parameters by \ac{MLB} significantly affects the \acp{KPI} of \ac{MRO}. For example, if \ac{MLB} modifies \ac{RET}, it directly impacts the handover boundary, potentially increasing the call drop rate. Direct and indirect conflicts between \acp{xApp} are readily identifiable; however, establishing implicit conflicts among \acp{xApp} is not feasible without live network simulation. Therefore, we are not considering any implicit conflicts among these four \acp{xApp} in this section.

The aforementioned conflicts, including the implicit conflict, can be resolved using our proposed \ac{QACM} method, as theoretically demonstrated in Section~\ref{sec:case_study}. However, practical validation requires these \acp{xApp} to be deployed in a \ac{Near-RT-RIC} and capable of predicting \acp{KPI}. We aim to validate the proposed \ac{QACM} method for all conflicts and demonstrate the results in future works. However, we perform a simulation study considering a direct conflict between \ac{ES} and \ac{CCO} \acp{xApp} over \ac{TXP} in the following subsection considering an actual \ac{RAN} scenario to showcase the performance of the proposed method. The simulation is performed without a \ac{RIC}, and \acp{xApp} are implemented as logic functions.

\begin{table*}[ht]
\centering
\caption{List of xApps with their respective KPIs}
\label{tab:kpi-xapps}
\begin{tabular}{|l|l|l|l|l|l|l|l|}
\hline
\textbf{xApp} & \textbf{KPI} & \textbf{QoS Range} & \textbf{Unit} & \multicolumn{4}{c|}{\textbf{Minimum QoS Threshold}} \\ \cline{5-8} 
 &  &  &  & \textbf{V2X} & \textbf{URLLC} & \textbf{mMTC} & \textbf{eMBB} \\ \hline
\multirow{2}{*}{CCO} & SINR & -10 to 30 dB & dB  \cite{alhammadi2023intelligent} & 20 dB & 15 dB & 10 dB & 15 dB \\ \cline{2-8} 
 & Downlink Throughput & 10 Mbps to 1 Gbps & Mbps/Gbps \cite{raca2020beyond} & 500 Mbps & 100 Mbps & 10 Mbps & 100 Mbps \\ \hline
\multirow{2}{*}{ES} & Energy Efficiency & 0.1 to 10 (bit/Joule) & bit/Joule \cite{wu2018energy} & 5 bit/Joule & 2 bit/Joule & 0.1-1 bit/Joule & 1-3 bit/Joule \\ \cline{2-8}
 & Power Consumption & 100 W to 1500 W & Watts \cite{shurdi20215g} & 1000 W & 500 W & 100 W & 500 W \\ \hline
\multirow{3}{*}{MRO} & Handover Success Rate & 90\% to 99\% & Percentage \cite{mu2014conflict} & 99\% & 98\% & 95\% & 97\% \\ \cline{2-8}
 & Call Drop Rate & 0\% to 2\% & Percentage \cite{mu2014conflict} & $\leq$ 0.5\% & $\leq$ 1\% & $\leq$ 2\% & $\leq$ 1.5\% \\ \cline{2-8}
 & Call Block Rate & 0\% to 2\% & Percentage \cite{mu2014conflict} & $\leq$ 0.5\% & $\leq$ 1\% & $\leq$ 2\% & $\leq$ 1.5\% \\ \hline
\multirow{2}{*}{MLB} & Traffic Load & 30\% to 80\% & Percentage & 60\% & 50\% & 30\% & 40\% \\ \cline{2-8}
 & Resource Utilization Rate & 20\% to 70\% & Percentage & 50\% & 40\% & 20\% & 30\% \\ \hline
\end{tabular}
\end{table*}

\begin{table}[ht]
\centering
\caption{Objectives of xApps based on their ICPs \cite{banerjee2021toward}}
\label{tab:xapps-objectives}
\begin{tabularx}{\columnwidth}{|l|X|X|}
\hline
\textbf{Name} & \textbf{ICPs}                  & \textbf{Objective}                          \\ \hline
MLB           & TTT, CIO, TXP, RET             & Minimize Load, Maximize Resource Use                               \\ \hline
CCO           & TXP, RET                       & Maximize Downlink Throughput, Minimize SINR                \\ \hline
ES            & TXP                            & Maximize Energy Efficiency, Minimize Power Use                     \\ \hline
MRO           & TXP, TTT, CIO, NL, HYS        & Maximize Handover Rate, Minimize Call Drops/Blocks \\ \hline
\end{tabularx}
\end{table}

\subsection{Validation Through Simulation} \label{sec:valid_sim}
To validate the proposed \ac{QACM} method, we used the MATLAB software for simulation and its 5G Toolbox with O-RAN 7.2 split \cite{arafat2024transformer}. The simulation involved two \acp{gNB} and ten \acp{UE}. The \acp{KPI} of interest, i.e., downlink throughput of \ac{CCO} \ac{xApp} and power consumption of \ac{ES} \ac{xApp}, were measured across different \ac{TXP} values. The simulation parameters included a \ac{RSRP} threshold of -110 dBm for handover, a frequency of 2.4 GHz, and a simulation time of 10 minutes with a time step of 100 ms. The QoS thresholds were set to 9.5 Gbps for throughput and 25 Wh for power consumption, ensuring that the system met the desired performance criteria. The input control parameters set during the simulations are: \ac{CIO} of 2 dB, \ac{HYS} of 0.5 dB, \ac{TTT} of 0.1 ms, \ac{RET} of 1.5 degree, and an adjustment interval of 1000 ms. The \acp{UE} are moving back-and-forth between the two \acp{gNB} with a randomly assigned velocity of 0 to 5m/s. Downlink throughput of \ac{CCO} needs to maximize and power consumption of \ac{ES} needs to minimize, therefore, we considered $\delta_{\rm{CCO}} = 0$ and $\delta_{\rm{ES}} = 1$ for the proposed \ac{QACM} method. 


\begin{figure}[!ht]
 \centering
 \vspace{-0.1in}
	\includegraphics[scale = 0.6]{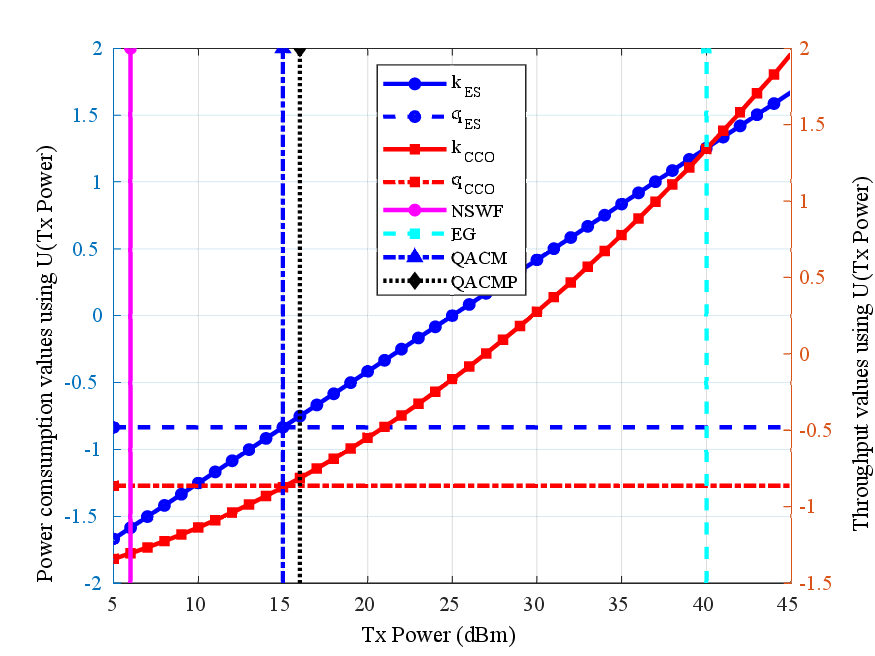}
	\vspace{-0.2in}
	\caption{Direct Conflict of  \ac{ES} and \ac{CCO} \acp{xApp} over \ac{TXP}.}
	\label{fig:simCon}
 \vspace{-0.1in}
\end{figure}

A direct conflict occurs during the simulation when the \ac{ES} \ac{xApp} sets \ac{TXP} to 6 dBm to minimize energy consumption in the network from its previous value of 40 dBm set by the \ac{CCO} \ac{xApp}. As a result, the downlink throughput \ac{KPI} belonging to \ac{CCO}, referred to as $k_{\rm{CCO}}$ in Fig.~\ref{fig:simCon}, experiences significant degradation below its threshold $q_{\rm{CCO}}$. To mitigate this conflict, we apply the \ac{QACM} conflict mitigation method and compare its performance with benchmark methods, similar to the case studies discussed in Section~\ref{sec:case_study}. The optimal configuration range is set to 6 to 40 dBm for finding the potential optimal value of \ac{TXP}. The \ac{QACM} and \ac{QACMP} methods, for non-priority and priority cases respectively, suggest an optimal \ac{TXP} of 15 dBm and 16 dBm, whereas the \ac{NSWF} and \ac{EG} methods suggest 6 dBm and 40 dBm, respectively. 

For non-priority cases with $w_{\text{ES}} = 0.5$ and $w_{\text{CCO}} = 0.5$, Fig.~\ref{fig:simCon} demonstrates that the value suggested by \ac{NSWF} for \ac{TXP} meets the power consumption threshold $q_{\text{ES}}$, where $k_{\text{ES}}$ is $142\%$ below its threshold. However, the throughput $k_{\text{CCO}}$ falls $39\%$ below its threshold $q_{\text{CCO}}$. In contrast, \ac{QACM} ensures that both \acp{xApp}' \acp{KPI} are as close as possible to their respective thresholds, with $k_{\text{ES}}$ exactly matching its threshold and $k_{\text{CCO}}$'s shortfall reduced from $39\%$ to $3\%$ relative to its threshold. For priority cases with $w_{\text{ES}} = 0.1$ and $w_{\text{CCO}} = 0.9$, the suggested value of \ac{TXP} by \ac{EG} results in power consumption increasing by $256\%$ and throughput by $200\%$ relative to their thresholds, creating an imbalance between the performances of these two \acp{xApp}. Conversely, \ac{QACMP} with a higher priority to the \ac{CCO} \ac{xApp} increases throughput by $3\%$ while only increasing power consumption by $5\%$, ensuring that both \acp{xApp} either meet their thresholds or stay as close as possible to them. Therefore, we can confirm that the proposed \ac{QACM} method outperforms other benchmark methods in \ac{QoS}-aware conflict mitigation and maximizes the number of \acp{xApp} meeting or closely approaching their respective \ac{QoS} thresholds.

\section{Limitation and Future Works}
\label{sec:limit_fut}
As discussed in Section~\ref{sec:application}, the primary focus of this work is the theoretical foundation of the \ac{QACM} method. While the \ac{KPI} prediction for \acp{xApp} plays a crucial role in our proposed approach, the current study employs a simplified \ac{ANN} model to illustrate the framework's potential. It is important to note that in a real \ac{RAN} environment, \ac{KPI} prediction is a complex process influenced not only by \acp{ICP} but also by the dynamic state of the network. The simplified \ac{KPI} prediction model used herein serves as a proof of concept for the underlying principles of the \ac{QACM} framework. Recognizing the need for a more comprehensive approach to \ac{KPI} prediction in actual network scenarios, we plan to conduct an in-depth investigation into this aspect in future research endeavors. Recent development \cite{tran2023ml} on \ac{KPI} prediction in \ac{RAN} shows the way to reach our vision.








\section{Conclusion}
\label{sec:conclusion}
In this article, we proposed the \ac{QACM} method that mitigates intra-component conflicts within the \ac{Near-RT-RIC} of Open \ac{RAN} architectures. By integrating elements of cooperative game theory, particularly \ac{NSWF} and \ac{EG} solutions, the \ac{QACM} method effectively balances conflicting parameters while upholding the individual \ac{QoS} requirements of \acp{xApp}. This approach not only enhances the flexibility and efficiency of network management but also  offers a standardized framework for conflict resolution among diverse network applications. The comparative analysis with benchmark methods in priority and non-priority scenarios further establishes the \ac{QACM} method's superiority in maintaining \ac{QoS} thresholds under conflicting conditions. However, it is crucial to recognize the simplicity of the \ac{KPI} prediction model used in this research may not comprehensively represent the complexity and dynamism of real-world \ac{RAN} environments. This limitation underscores the need for more comprehensive \ac{KPI} prediction models and the integration of the conflict mitigation framework introduced in this paper within the testbed environments that we aim to investigate in future. Also, we envisioned the concept of \ac{CS} xApp for conflict supervision which is an integral part of the \ac{QACM} based conflict mitigation system. We aim to dive deeper into the implementation of each of these components.
Our future research will focus on bridging the gap between theoretical models and practical applications. In doing so, we aim to reinforce the role of QACM as an essential component in Open \ac{RAN}, ensuring optimized performance and enhanced user experience.



\section*{Acknowledgment}
Funding for this research is partially provided by the European Union's Horizon Europe research and innovation program through the Marie Skłodowska-Curie SE grant under agreement number RE-ROUTE No 101086343.

\ifCLASSOPTIONcaptionsoff
  \newpage
\fi



%
\bibliographystyle{ieeetr}
\bibliography{references}

%








\begin{IEEEbiography}[{\includegraphics[width=1in,height=1.25in,clip,keepaspectratio]{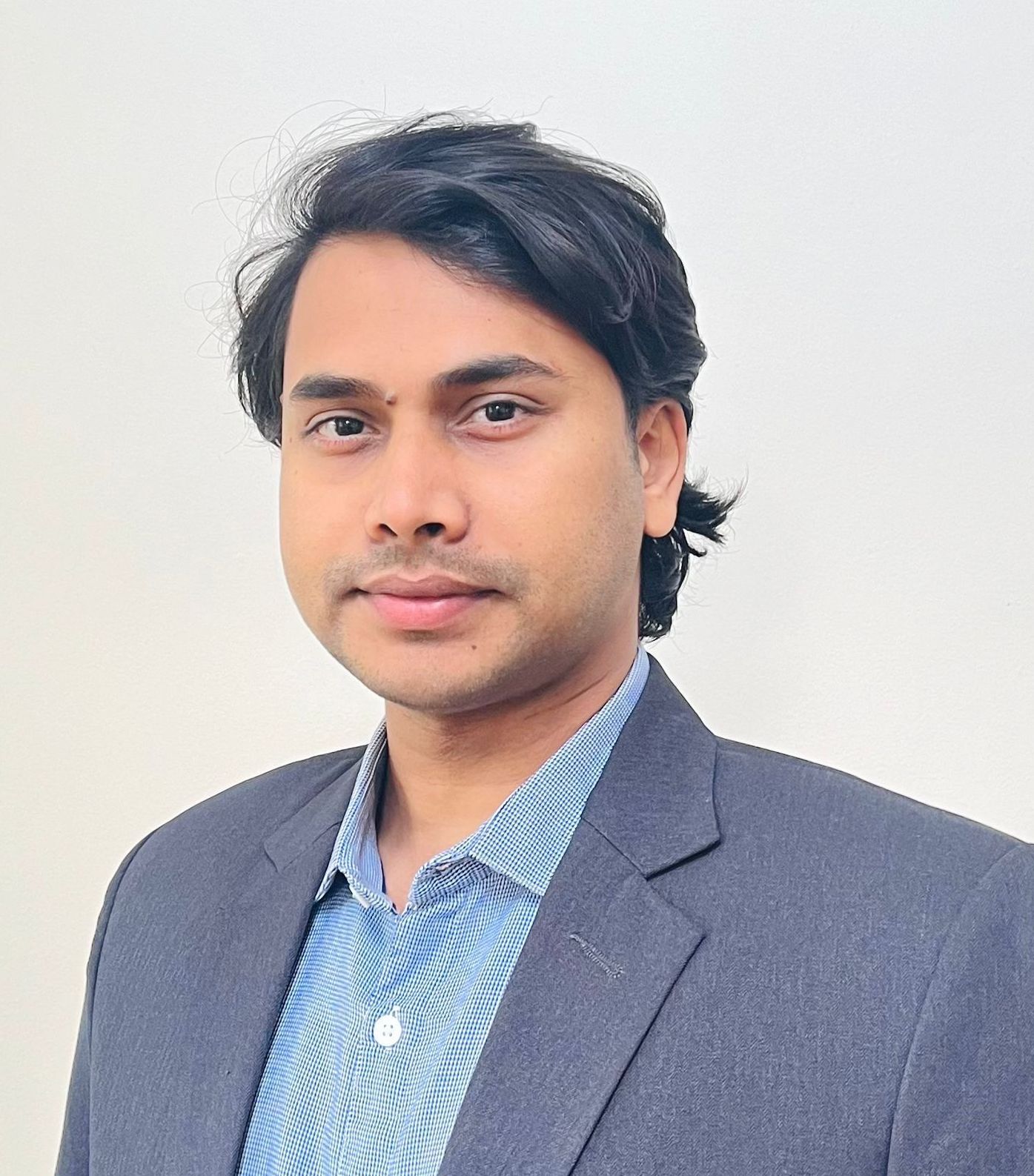}}]
{Abdul Wadud} is a Ph.D. Researcher at the UCD School of Computer Science, Ireland, and a Research Associate at the Bangladesh Institute of Governance and Management (BIGM), Dhaka. He received an MSc degree in Computer Science from South Asian University (SAU), India, in July 2020. He was awarded the prestigious President Scholarship and the SAU Special Scholarship at SAU. He has published research articles in top-tier journals like \textit{IEEE/ACM Transactions on Networking}, \textit{IEEE Transactions on Network and Service Management}, and \textit{IEEE Journal on Selected Areas in Communications}. His research interests include open RAN, wireless mobile networks, optical networks, optimization, and AI/ML for communication. He is a graduate student member of IEEE.
\end{IEEEbiography}

\begin{IEEEbiography}[{\includegraphics[width=1in,height=1.25in,clip,keepaspectratio]{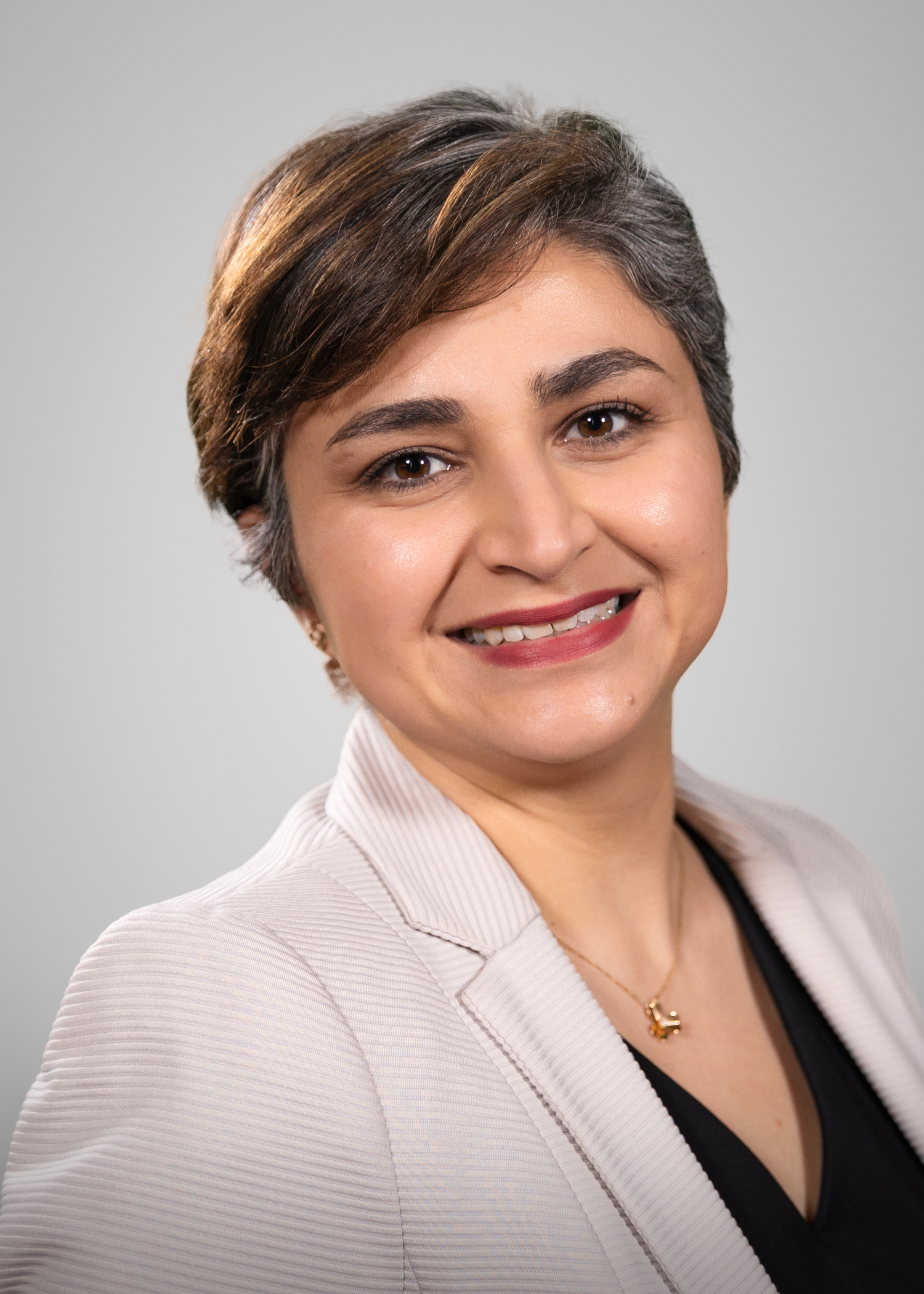}}]
{Dr. Fatemeh Golpayegani} is an Assistant Professor in the school of Computer Science, University College Dublin, and a Young Academy Ireland member.  She  received her PhD from Trinity College Dublin in 2018 and joined UCD in 2019. She leads the Multiagent and sustainable solutions research group, including 3 postdoctoral researchers, and 6 PhD students. Within her research group the main focus is on developing AI-powered decision making algorithms, and optimisation algorithms for complex systems and environments. Her research is applied in several fields including Intelligent Transport Systems, Power Systems and Bio systems. She has secured over 2 million euros in funding through international and national sources. Specifically she is coordinating an MSCA SE project, RE-ROUTE funded under Horizon Europe. She is the co-PI of another EU project, Augmented CCAM. She is a funded investigator of SFI research centres, including BiOrbic, CONNECT, I-Form, and funded supervisor in SFI research and training centres ML-labs and D-real.
\end{IEEEbiography}

\begin{IEEEbiography}[{\includegraphics[width=1in,height=1.25in,clip,keepaspectratio]{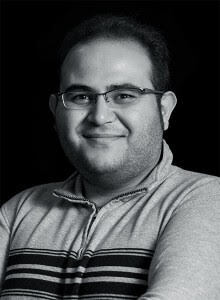}}]
{Dr. Nima Afraz} is a tenured Assistant Professor in the School of Computer Science at University College Dublin. He earned his PhD in Computer Science from Trinity College Dublin, Ireland, in 2020. Nima is a funded investigator with the CONNECT Research Centre, where his research is focused on Open Radio Access Networks, blockchain applications in telecommunications, network economics, and network virtualization. He is a recipient of the Government of Ireland Postdoctoral Fellowship and has worked as a postdoctoral fellow addressing challenges related to the adoption of blockchain technology in telecommunications. Nima is a coordinator of an EU MSCA Project RE-ROUTE and has made significant contributions to the Broadband Forum standard TR-402, and the ETSI GS PDL 022 specification. Additionally, he serves as the vice-chair of the Linux Foundation's Hyperledger Telecom Special Interest Group.
\end{IEEEbiography}

\end{document}

%% file: Acronyms.tex
\begin{acronym}[BEREC]
\acro{10G-EPON}{10 Gigabit EPON}
\acro{3GPP}{3rd Generation Partnership Project}
\acro{5G}{Fifth Generation}
\acro{6G}{Sixth Generation}
\acro{A-CPI}{A Controller Plane Interface}
\acro{A-RoF}{Analog Radio-over-Fiber}
\acro{ABNO}{Application-Based Network Operations}
\acro{ADC}{Analog-to-Digital Converter}
\acro{AE}{Allocative Efficiency}
\acro{amc}{adaptive modulation and coding}
\acro{AON}{Active Optical Network}
\acro{AI}{Artificial Intelligence}
\acro{ML}{Machine Learning}
\acro{API}{Application Programming Interface}
\acro{AWG}{Arrayed Waveguide Grating}
\acro{B2B}{Business to Business}
\acro{B2C}{Business to Consumer}
\acro{B-RAS}{Broadband Remote Access Servers}
\acro{BaaS}{Blockchain as a Service}
\acro{BB}{Budget Balance}
\acro{BBF}{BroadBand Forum}
\acro{BBU}{Base Band Unit}
\acro{BER}{Bit Error Rate}
\acro{BEREC}{Body of European Regulators for Electronic Communications}
\acro{BF}{beamforming}
\acro{BFT}{Byzentine fault tolerant}
\acro{BGP}{Border Gateway Protocol}
\acro{BMap}{Bandwidth Map}
\acro{BPM}{Business Process Management}
\acro{BS}{Base Station}
\acro{BSC}{Base Station Controller}
\acro{BtB}{Back to Back}
\acro{BW}{Bandwidth}
\acro{C2C}{Consumer to Consumer}
\acro{C-RAN}{Cloud Radio Access Network}
\acro{C-RoFN}{Cloud-based Radio over optical Fiber Network}
\acro{CA}{Certificate Authority}
\acro{CapEx}{Capital Expenditure}
\acro{CDMA}{code division multiple access}
\acro{CDR}{Clock Data Recovery}
\acro{CFO}{carrier frequency offset}
\acro{CIR}{Committed Information Rate}
\acro{CO}{Central Office}
\acro{CoMP}{Coordinated Multipoint}
\acro{COP}{Control Orchestration Protocol}
\acro{CORD}{Central Office Rearchitected as a Data Centre}
\acro{CPE}{Customer Premises Equipment}
\acro{CPRI}{Common Public Radio Interface}
\acro{CPU}{Central Processing Unit}
\acro{CMC}{Conflict Mitigation Controller}
\acro{CMS}{Conflict Management System}
\acro{CQI}{channel quality indicator}
\acro{CR}{cognitive radio}
\acro{CS}{Central Station}
\acro{CSI}{channel state information}
\acro{D2D}{Device to Device}
\acro{D-CPI}{D Controller Plane Interface}
\acro{D-RoF}{Digital Radio-over-Fiber}
\acro{DAC}{Digital-to-Analog Converter}
\acro{DAS}{distributed antenna system}
\acro{DAS}{Distributed Antenna Systems}
\acro{DBA}{Dynamic Bandwidth Allocation}
\acro{DBRu}{Dynamic Bandwidth Report upstream}
\acro{DC}{Direct Current}
\acro{DL}{Downlink}
\acro{DLT}{Distributed Ledger Technology}
\acro{DMT}{Discrete Multitone}
\acro{DNS}{Domain Name Server}
\acro{DPDK}{Data Plane Development Kit}
\acro{DSB}{Double Side Band}
\acro{DSL}{Digital Subscriber Line}
\acro{DSLAM}{Digital Subscriber Line Access Multiplexer}
\acro{DSP}{Digital Signal Processing}
\acro{DSS}{Distributed Synchronization Service}
\acro{DU}{Distributed Unit}
\acro{DUE}{D2D user equipment}
\acro{DWDM}{Dense Wave Division Multiplexing}
\acro{E-CORD}{Enterprise CORD}
\acro{EAL}{Environment Abstraction Layer}
\acro{EDF}{Erbium-Doped Fiber}
\acro{EFM}{Ethernet in the First Mile}
\acro{EMBS}{Elastic Mobile Broadband Service}
\acro{EON}{Elastic Optical Network}
\acro{EPC}{Evolved Packet Core}
\acro{EPON}{Ethernet Passive Optical Network}
\acro{eMBB}{enhanced Mobile Broadband}
\acro{ETSI}{European Telecommunications Standards Institute}
\acro{EVM}{Error Vector Magnitude}
\acro{FANS}{Fixed Access Network Sharing}
\acro{FASA}{Flexible Access System Architecture}
\acro{FBMC}{filter bank multi-carrier}
\acro{FBMC}{Filterbank Multicarrier}
\acro{FCC}{Federal Communications Commission}
\acro{FDD}{Frequency Division Duplex}
\acro{FDMA}{frequency-division multiple access}
\acro{FEC}{Forward Error Correction}
\acro{FFR}{Fractional Frequency Reuse}
\acro{FFT}{Fast Fourier Transform}
\acro{FPGA}{Field-Programmable Gate Array}
\acro{FSO}{Free-Space-Optics}
\acro{FSR}{Free Spectral Range}
\acro{FTN}{faster-than-nyquist}
\acro{FTTC}{Fiber-to-the-Curb}
\acro{FTTH}{Fiber-to-the-Home}
\acro{FTTx}{Fiber-to-the-x}
\acro{G.Fast}{Fast Access to Subscriber Terminals}
\acro{GEM}{G-PON encapsulation method}
\acro{GFDM}{Generalized Frequency Division Multiplexing}
\acro{GMPLS}{Generalized Multi-Protocol Label Switching}
\acro{GPON}{Gigabit Passive Optical Network}
\acro{GPP}{General Purpose Processor}
\acro{GPRS}{General Packet Radio Service}
\acro{GSM}{Global System for Mobile Communications}
\acro{GTP}{GPRS Tunneling Protocol}
\acro{GUI}{Graphical User Interface}
\acro{HA}{Hardware Accelerator}
\acro{HARQ}{Hybrid-Automatic Repeat Request}
\acro{HD}{half duplex}
\acro{HetNet}{heterogeneous networks}
\acro{IaaS}{Infrastructure as a Service}
\acro{I-CPI}{I Controller Plane Interface}
\acro{I2RS}{Interface 2 Routing System}
\acro{IA}{Interference Alignment}
\acro{IAM}{Identity and Access Management}
\acro{IBFD}{in-band full duplex}
\acro{IC}{Incentive Compatibility}
\acro{ICIC}{inter-cell interference coordination}
\acro{ICT}{information and communications technology}
\acro{IEEE}{Institute of Electrical and Electronics Engineers}
\acro{IETF}{Internet Engineering Task Force}
\acro{IF}{Intermediate Frequency}
\acro{IFFT}{inverse FFT}
\acro{ITS}{Intelligent Transport Systems}
\acro{InP}{Infrastructure Provider}
\acro{IoT}{Internet of Things}
\acro{IP}{Internet Protocol}
\acro{IQ}{In-phase/Quadrature}
\acro{IR}{Individual Rationality}
\acro{IRC}{Interference Rejection Combining}
\acro{ISI}{inter-symbol interference}
\acro{ISP}{Internet Service Provider}
\acro{BFT}{Byzantine-Fault-Tolerant}
\acro{IV}{Intelligent Vehicle}
\acro{ITU}{International Telecommunication Union}
\acro{KPI}{Key Performance Indicator}
\acro{KVM}{Kernel-based Virtual Machine}
\acro{L2}{Layer-2}
\acro{L3}{Layer-3}
\acro{LAN}{Local Area Network}
\acro{LO}{Local Oscillator}
\acro{LOS}{Line Of Sight}
\acro{LR-PON}{Long Reach PON}
\acro{LSP-DB}{Label Switched Path Database}
\acro{LTE-A}{Long Term Evolution Advanced}
\acro{LTE}{Long Term Evolution}
\acro{M-CORD}{Mobile CORD}
\acro{M-MIMO}{massive MIMO}
\acro{M2M}{machine-to-machine}
\acro{MAC}{Medium Access Control}
\acro{ME}{Merging Engine}
\acro{MIMO}{multiple input multiple output}
\acro{MME}{Mobility Management Entity}
\acro{mmWave}{millimeter wave}
\acro{MNO}{mobile network operator}
\acro{MPLS}{Multiprotocol Label Switching}
\acro{MRC}{Maximum Ratio Combining}
\acro{MSC}{Mobile Switching Centre}
\acro{MSP}{Membership Service Providers}
\acro{MSR}{Multi-Stratum Resources}
\acro{MTC}{machine type communication}
\acro{mMTC}{massive Machine Type Communications}
\acro{MVNO}{Mobile Virtual Network Operator}
\acro{MVNP}{mobile virtual network provider}
\acro{MZI}{Mach-Zehnder Interferometer}
\acro{MZM}{Mach-Zehnder Modulator}
\acro{NE}{Network Element}
\acro{NFV}{Network Function Virtualization}
\acro{NFVaaS}{Network Function Virtualization as a Service}
\acro{NG-PON2}{Next-Generation Passive Optical Network 2}
\acro{NGA}{Next Generation Access}
\acro{NLOS}{None Line Of Sight}
\acro{NMS}{Network Management System}
\acro{NRZ}{Non Return-to-Zero}
\acro{NTT}{Nippon Telegraph and Telephone}
\acro{OBPF}{Optical BandPass Filter}
\acro{OBSAI}{Open Base Station Architecture Initiative}
\acro{ODN}{Optical Distribution Network}
\acro{OFC}{Optical Frequency Comb}
\acro{OFDM}{orthogonal frequency-division multiplexing}
\acro{OFDMA}{orthogonal frequency-division multiple access}
\acro{OLO}{Other Licensed Operator}
\acro{OLT}{Optical Line Terminal}
\acro{ONF}{Open Networking Foundation}
\acro{ONU}{Optical Network Unit}
\acro{OOB}{out-of-band}
\acro{OOK}{On-off Keying}
\acro{OpenCord}{Central Office Re-Architected as a Data Center}
\acro{OpEx}{Operating Expenditure}
\acro{OSI}{Open Systems Interconnection}
\acro{OSS}{Operations Support Systems}
\acro{OTT}{Over-the-Top}
\acro{OXM}{OpenFlow Extensible Match}
\acro{P2MP}{Ethernet over point-to-multipoint }
\acro{P2P}{Point-to-Point }
\acro{PAM}{Pulse Amplitude Modulation}
\acro{PAPR}{peak-to-average power rating}
\acro{PAPR}{Peak-to-Average Power Ratio}
\acro{PBMA}{Priority Based Merging Algorithm}
\acro{PCE}{Path Computation Elements}
\acro{PCEP}{Path Computation Element Protocol}
\acro{PCF}{Photonic Crystal Fiber}
\acro{PD}{Photodiode}
\acro{PDCP}{RD Control Protocol}
\acro{PDF}{probability distribution function}
\acro{PGW}{Packet Gateway}
\acro{PHY}{physical Layer}
\acro{PIR}{Peak Information Rate}
\acro{PMD}{Polarization Division Multiplexing}
\acro{PON}{Passive Optical Network}
\acro{POTS}{Plain Old Telephone Service}
\acro{PoET}{Proof of Elapsed Time}
\acro{PoW}{Proof of Work}
\acro{PoS}{Proof of Stake}
\acro{PTP}{Precision Time Protocol}
\acro{PWM}{Pulse Width Modulation}
\acro{QAM}{Quadrature Amplitude Modulation}
\acro{QoE}{Quality of Experience}
\acro{QoS}{Quality of Service}
\acro{QPSK}{Quadrature Phase Shift Keying}
\acro{R-CORD}{Residential CORD}
\acro{RA}{Resource Allocation}
\acro{RAN}{Radio Access Network}
\acro{RAT}{Radio Access Technology}
\acro{RB}{resource block}
\acro{RFIC}{radio frequency integrated circuit}
\acro{RIC}{RAN Intelligent Controller}
\acro{RLC}{Radio Link Control}
\acro{RN}{Remote Node}
\acro{ROADM}{Reconfigurable Optical Add Drop Multiplexer}
\acro{RoF}{Radio-over-Fiber}
\acro{RRC}{Radio Resource Control}
\acro{RRH}{Remote Radio Head}
\acro{RRPH}{Remote Radio and PHY Head}
\acro{RRU}{Remote Radio Unit}
\acro{RSOA}{Reflective Semiconductor Optical Amplifier}
\acro{RU}{Remote Unit}
\acro{Rx}{receiver}
\acro{SC-FDMA}{single carrier frequency division multiple access}
\acro{SCM}{single carrier modulation}
\acro{SD-RAN}{Software Defined Radio Access Network}
\acro{SDAN}{Software Defined Access Network}
\acro{SDMA}{Semi-Distributed Mobility Anchoring}
\acro{SDN}{Software Defined Networking}
\acro{SDR}{Software Defined Radio}
\acro{SFBD}{Single Fiber Bi-Direction}
\acro{SGW}{Serving Gateway}
\acro{SI}{self-interference}
\acro{SIC}{self-interference cancellation}
\acro{SIMO}{Single Input Multiple Output}
\acro{SLA}{Service Level Agreement}
\acro{SMF}{Single Mode Fiber}
\acro{SNMP}{Simple Network Management Protocol}
\acro{SNR}{Signal-to-Noise Ratio}
\acro{SP}{Service Providers}
\acro{Split-PHY}{Split Physical Layer}
\acro{SRI-OV}{Single Root Input/Output Virtualization}
\acro{SU}{secondary user}
\acro{SUT}{System Under Test}
\acro{T-CONT}{Transmission Container}
\acro{TC}{Transmission Convergence}
\acro{TCO}{Total Cost of Ownership}
\acro{TD-LTE}{Time Division LTE}
\acro{TDD}{Time Division Duplex}
\acro{TDM}{Time Division Multiplexing}
\acro{TDMA}{Time Division Multiple Access}
\acro{TED}{Traffic Engineering Database}
\acro{TEID}{Tunnel endpoint identifier}
\acro{TLS}{Transport Layer Security}
\acro{TPS}{Transactions per Second}
\acro{TTI}{transmission time interval}
\acro{TWDM}{Time and Wavelength Division Multiplexing}
\acro{Tx}{transmitter}
\acro{UD-CRAN}{Ultra-Dense Cloud Radio Access Network}
\acro{UDP}{User Datagram Protocol}
\acro{UE}{User Equipment}
\acro{UF-OFDM}{Universally Filtered OFDM}
\acro{UFMC}{Universally Filtered Multicarrier}
\acro{UL}{Uplink}
\acro{UMTS}{Universal Mobile Telecommunications System}
\acro{URLLC}{Ultra-Reliable Low-Latency Communication}
\acro{USRP}{Universal Software Radio Peripheral}
\acro{V2I}{Vehicle to Infrastructure}
\acro{V2V}{Vehicle to Vehicle}
\acro{vBBU}{virtualized BBU}
\acro{vBMap}{Virtual Bandwidth Map}
\acro{vBS}{virtual Base station}
\acro{VCG}{Vickery-Clarke-Groves}
\acro{vCPE}{virtual CPE}
\acro{vCPU}{Virtual Central Processing Unit}
\acro{vDBA}{virtual DBA}
\acro{VDSL2}{Very-high-bit-rate digital subscriber line 2}
\acro{VLAN}{Virtual Local Area Network}
\acro{VM}{Virtual Machine}
\acro{VNF}{Virtual Network Function}
\acro{VNI}{visual networking index}
\acro{VNO}{Virtual Network Operator}
\acro{VNTM}{Virtual Network Topology Manager}
\acro{vOLT}{virtual OLT}
\acro{VPE}{virtual Provider Edge}
\acro{VPN}{Virtual Private Network}
\acro{V2X}{Vehicle to Everything}
\acro{VTN}{Virtual Tenant Network}
\acro{VULA}{Virtual Unblunded Local Access}
\acro{WAN}{Wide Area Network}
\acro{WAP}{Wireless Access Point}
\acro{WCDMA}{wide-band code division multiple access}
\acro{WDM-PON}{Wavelength Division Multiplexing - Passive Optical Network}
\acro{WDM}{Wavelength Division Multiplexing}
\acro{WiMAX}{Worldwide Interoperability for Microwave Access}
\acro{WRPR}{Wired-to-RF Power Ratio}
\acro{XG-PON}{10 Gigabit PON}
\acro{XGEM}{XG-PON encapsulation method}
\acro{XOS}{XaaS Operating System}

\acro{RIC}{RAN Intelligent Controller}
\acro{Near-RT-RIC}{Near Real Time RAN Intelligent Controller}
\acro{Non-RT-RIC}{Non Real Time RAN Intelligent Controller}
\acro{xApp}{Extended Application}
\acro{rApp}{Remote Application}
\acro{CMS}{Conflict Mitigation System}
\acro{EG}{Eisenberg-Galle}
\acro{NSWF}{Nash's Social Welfare Function} 
\acro{CMC}{Conflict Mitigation Controller}
\acro{CDC}{Conflict Detection Controller}
\acro{PMon}{Performance Monitoring}
\acro{RCP}{Recently Changed Parameter}
\acro{PGD}{Parameter Group Definition}
\acro{RCPG}{Recently Changed Parameter Group}
\acro{DCKD}{Decision Correlated with KPI Degradation} 
\acro{KDO}{KPI Degradation Occurrences}
\acro{MNO}{Mobile Network Operator}
\acro{SDL}{Shared Data Layer}
\acro{SON}{Self Organising Network}
\acro{SF}{SON Function}
\acro{AM}{Arithmetic Mean}
\acro{PKR}{Parameter and KPI Ranges}
\acro{KPI}{Key Performance Indicator}
\acro{ICP}{Input Control Parameter}
\acro{RF}{Radio Frequency}
\acro{C-RAN}{Cloud Radio Access Network}
\acro{MARL}{Multi Agent Reinforcement Learning}
\acro{CCO}{Capacity and Coverage Optimization}
\acro{ES}{Energy Saving}
\acro{MRO}{Mobility Robustness Optimization}
\acro{MLB}{Mobility Load Balancing}
\acro{V-RAN}{Virtualized Radio Access Network}
\acro{O-RAN}{Open Radio Access Network}
\acro{D-RAN}{Distributed Radio Access Network}
\acro{5GB}{5G and Beyond}
\acro{O-RU}{Open Radio Unit}
\acro{O-DU}{Open Distribution Unit}
\acro{O-CU}{Open Centralised Unit}
\acro{O-CU-CP}{Open Centralised Unit - Control Plane}
\acro{O-CU-UP}{Open Centralised Unit - User Plane}
\acro{D-SON}{Distributed SON}
\acro{C-SON}{Centralised SON}
\acro{CIO}{Cell Individual Offset}
\acro{CBR}{Call Block Ratio}
\acro{HOR}{Handover Ratio}
\acro{CAN}{Cognitive Autonomous Network}
\acro{CF}{Cognitive Function}
\acro{QACM}{QoS-Aware Conflict Mitigation}
\acro{ANN}{Artificial Neural Network}
\acro{PR}{Polynomial Regression}
\acro{MSE}{Mean Squared Error}
\acro{EVS}{Explained Variance Score}
\acro{CS}{Conflict Supervision}
\acro{TXP}{Tx Power}
\acro{TTT}{Time-To-Trigger}
\acro{CIO}{Cell Individual Offset}
\acro{NL}{Neighbour List}
\acro{RET}{Radio Electrical Tilt}
\acro{HYS}{Handover Hysteresis}
\acro{QACMP}{QACM Priority}
\acro{DRL}{Deep Reinforcement Learning}
\acro{gNB}{Next-generation NodeB}
\acro{RSRP}{Reference Signal Received Power}
\end{acronym}

%% file: manuscript.bbl
\begin{thebibliography}{10}

\bibitem{wadud2023conflict}
A.~Wadud, F.~Golpayegani, and N.~Afraz, ``Conflict management in the
  near-rt-ric of open ran: A game theoretic approach,'' {\em arXiv preprint
  arXiv:2311.13389}, 2023.

\bibitem{polese2022understanding}
M.~Polese, L.~Bonati, S.~D'Oro, S.~Basagni, and T.~Melodia, ``Understanding
  o-ran: Architecture, interfaces, algorithms, security, and research
  challenges,'' {\em arXiv preprint arXiv:2202.01032}, 2022.

\bibitem{rimedo_labs}
M.~Dryjański, ``Traffic management for v2x use cases in o-ran.''
  \url{https://rimedolabs.com/blog/traffic-management-for-v2x-use-cases-in-o-ran/}.
\newblock Accessed on: 09-11-2023.

\bibitem{b3}
{O-RAN Alliance}, ``O-ran use cases analysis report,'' Report v09.00, O-RAN
  Alliance, Oct. 2022.

\bibitem{ric_oran_alliance}
O.-R. Alliance, ``O-ran working group 3 (near-real-time ran intelligent
  controller and e2 interface workgroup), near-rt ric architecture,'' {\em
  O-RAN.WG3.RICARCH-R003-v04.00}, Last Accessed [March 2023].

\bibitem{adamczyk2023conflict}
C.~Adamczyk and A.~Kliks, ``Conflict mitigation framework and conflict
  detection in o-ran near-rt ric,'' {\em IEEE Communications Magazine}, 2023.

\bibitem{zhang2022team}
H.~Zhang, H.~Zhou, and M.~Erol-Kantarci, ``Team learning-based resource
  allocation for open radio access network (o-ran),'' in {\em ICC 2022-IEEE
  International Conference on Communications}, pp.~4938--4943, IEEE, 2022.

\bibitem{del2024pacifista}
P.~B. del Prever, S.~D'Oro, L.~Bonati, M.~Polese, M.~Tsampazi, H.~Lehmann, and
  T.~Melodia, ``Pacifista: Conflict evaluation and management in open ran,''
  {\em arXiv preprint arXiv:2405.04395}, 2024.

\bibitem{adamczyk2023challenges}
C.~Adamczyk, ``Challenges for conflict mitigation in o-ran’s ran intelligent
  controllers,'' in {\em 2023 International Conference on Software,
  Telecommunications and Computer Networks (SoftCOM)}, pp.~1--6, IEEE, 2023.

\bibitem{dryjanski2021toward}
M.~Dryja{\'n}ski, {\L}.~Ku{\l}acz, and A.~Kliks, ``Toward modular and flexible
  open ran implementations in 6g networks: Traffic steering use case and o-ran
  xapps,'' {\em Sensors}, vol.~21, no.~24, p.~8173, 2021.

\bibitem{Mavenir_RIC}
{Mavenir}, ``{RIC as the Next Generation SON for Open RAN and More}.''
  \url{https://www.mavenir.com/resources/ric-as-the-next-generation-son-for-open-ran-and-more/}.

\bibitem{SNS_SON}
{SNS Telecom \& IT}, ``{SON (Self-Organizing Networks) in the 5G \& Open RAN
  Era: 2022 – 2030 – Opportunities, Challenges, Strategies \& Forecasts}.''
  \url{https://www.snstelecom.com/son}.
\newblock Accessed: 2024-01-01.

\bibitem{liu2010conflict}
Z.~Liu, P.~Hong, K.~Xue, and M.~Peng, ``Conflict avoidance between mobility
  robustness optimization and mobility load balancing,'' in {\em 2010 IEEE
  Global Telecommunications Conference GLOBECOM 2010}, pp.~1--5, IEEE, 2010.

\bibitem{mu2014conflict}
P.~Mu, R.~Barco, S.~Fortes, {\em et~al.}, ``Conflict resolution between load
  balancing and handover optimization in lte networks,'' {\em IEEE
  Communications Letters}, vol.~18, no.~10, pp.~1795--1798, 2014.

\bibitem{huang2018conflict}
M.~Huang and J.~Chen, ``A conflict avoidance scheme between mobility load
  balancing and mobility robustness optimization in self-organizing networks,''
  {\em Wireless Networks}, vol.~24, pp.~271--281, 2018.

\bibitem{anubhab2020conflict}
A.~Banerjee, S.~S. Mwanje, and G.~Carle, ``Game theoretic conflict resolution
  mechanism for cognitive autonomous networks,'' in {\em 2020 International
  Symposium on Performance Evaluation of Computer and Telecommunication Systems
  (SPECTS)}, pp.~1--8, 2020.

\bibitem{banerjee2021toward}
A.~Banerjee, S.~S. Mwanje, and G.~Carle, ``Toward control and coordination in
  cognitive autonomous networks,'' {\em IEEE Transactions on Network and
  Service Management}, vol.~19, no.~1, pp.~49--60, 2021.

\bibitem{datagithub}
``Qacm repository.'' \url{https://github.com/dewanwadud1/QACM}.
\newblock Last Updated: [Jan 15, 2024].

\bibitem{colan2013and}
S.~D. Colan, ``The why and how of z scores,'' {\em Journal of the American
  Society of Echocardiography}, vol.~26, no.~1, pp.~38--40, 2013.

\bibitem{tran2023ml}
N.~P. Tran, O.~Delgado, B.~Jaumard, and F.~Bishay, ``Ml kpi prediction in 5g
  and b5g networks,'' in {\em 2023 Joint European Conference on Networks and
  Communications \& 6G Summit (EuCNC/6G Summit)}, pp.~502--507, IEEE, 2023.

\bibitem{alhammadi2023intelligent}
A.~Alhammadi, W.~H. Hassan, A.~A. El-Saleh, I.~Shayea, H.~Mohamad, and W.~K.
  Saad, ``Intelligent coordinated self-optimizing handover scheme for 4g/5g
  heterogeneous networks,'' {\em ICT Express}, vol.~9, no.~2, pp.~276--281,
  2023.

\bibitem{raca2020beyond}
D.~Raca, D.~Leahy, C.~J. Sreenan, and J.~J. Quinlan, ``Beyond throughput, the
  next generation: a 5g dataset with channel and context metrics,'' in {\em
  Proceedings of the 11th ACM multimedia systems conference}, pp.~303--308,
  2020.

\bibitem{wu2018energy}
J.~Wu, R.~Tan, and M.~Wang, ``Energy-efficient multipath tcp for
  quality-guaranteed video over heterogeneous wireless networks,'' {\em IEEE
  Transactions on Multimedia}, vol.~21, no.~6, pp.~1593--1608, 2018.

\bibitem{shurdi20215g}
O.~Shurdi, L.~Ruci, A.~Biberaj, and G.~Mesi, ``5g energy efficiency overview,''
  {\em European Scientific Journal}, vol.~17, no.~3, pp.~315--27, 2021.

\bibitem{arafat2024transformer}
M.~Arafat~Habib, P.~E. Iturria-Rivera, Y.~Ozcan, M.~Elsayed, M.~Bavand,
  R.~Gaigalas, and M.~Erol-Kantarci, ``Transformer-based wireless traffic
  prediction and network optimization in o-ran,'' {\em arXiv e-prints},
  pp.~arXiv--2403, 2024.

\end{thebibliography}
